\documentclass[11pt,preprint]{aastex}
\usepackage{epsfig,color}
\usepackage{mathrsfs}

\shorttitle{Intra-day optical variability in S5 0716+714 and 3C 273} \shortauthors{Liu et al.}

\begin{document}

\title{Search for Intra-day Optical Variability in $\gamma$-ray--loud Blazars S5 0716+714 and 3C 273}

\author{H. T. Liu\altaffilmark{1,2,3}$^{\bigstar}$, Hai Cheng Feng\altaffilmark{1,2,4}, Y. X. Xin\altaffilmark{1,2,3,4}, J. M. Bai\altaffilmark{1,2,3}, S. K. Li\altaffilmark{1,2,3}, and Fang Wang\altaffilmark{5}}

\altaffiltext{1}{Yunnan Observatories, Chinese Academy of Sciences,
Kunming 650011, Yunnan, P. R. China.}

\altaffiltext{2}{Key Laboratory for the Structure and Evolution of
Celestial Objects, Chinese Academy of Sciences, Kunming 650011,
Yunnan, China}

\altaffiltext{3}{Center for Astronomical Mega-Science, Chinese Academy of Sciences, 20A Datun Road, Chaoyang District, Beijing, 100012, P. R. China}

\altaffiltext{4}{University of Chinese Academy of Sciences, Beijing 100049, P. R. China}

\altaffiltext{5}{School of Physics and Space Science, China West Normal University, Nanchong, 637009, P. R. China}

\altaffiltext{$^{\bigstar}$}{Corresponding author: H. T. Liu, htliu@ynao.ac.cn}

\begin{abstract}
We present the photometric observations of blazars S5 0716+714 and 3C 273 with high temporal resolution (30--60s) in the $I$ or $R$ bands. The observations were performed with a 1.02 m optical telescope from 2007 March 07 to 2012 May 16. The $F$-test, one-way analysis of variance (ANOVA) test, and $z$-transformed discrete correlation function (ZDCF) cross-correlation zero lag test are used to search for intra-day variability (IDV). Four and five reliable IDVs survive three tests for S5 0716+714 and 3C 273, respectively. IDVs are found for S5 0716+714 and 3C 273. A flare on 2008 May 08 has $\Delta I \approx$ 0.06$\pm$0.01 mag in a duration of 0.54 hr for S5 0716+714. A flare on 2011 May 10 shows $\Delta R \approx$ 0.05$\pm$0.01 mag in a duration of 0.40 hr for 3C 273. Sharp dips appear on 2011 May 9 for 3C 273, and show $\Delta R \approx$ 0.05$\pm$0.01 mag. Under the assumptions that the IDV is tightly connected to black hole mass, $M_{\bullet}$, and that the flare durations are representative of the minimum characteristic timescales, we can estimate upper bounds to $M_{\bullet}$. In the case of the Kerr black holes, $M_{\bullet} \la 10^{8.91} M_{\odot}$ and $M_{\bullet} \la 10^{9.02} M_{\odot}$ are given for S5 0716+714 and 3C 273, respectively. These mass measurements are consistent with those measurements reported in the literatures. Also, we discuss the origins of optical variations found in this work.

\end{abstract}


\section{INTRODUCTION}

Blazars are a special subclass of radio-loud active galactic nuclei (AGNs) and show significant properties, such as rapid and strong variability from radio to $\gamma$-ray bands, high and variable polarization, prominent nonthermal emission,
etc \citep[e.g.,][]{UP95}. These significant properties are mostly generated by a relativistic jet with a small viewing angle $\lesssim$ 10$^{\circ}$ \citep[e.g.,][]{BK79,UP95}. Spectral energy distributions (SEDs) of blazars generally show
a double-peak profile in broadband continuum from radio to $\gamma$-ray bands. Broadband observations show that the
low energy peak is from infrared (IR)--optical--ultraviolet (UV) to soft X-ray bands, and the high energy peak is in the MeV--GeV--TeV $\gamma$-ray regime \citep[e.g.,][]{Gh98,Ab10b}. The usual classification between flat spectrum radio qusasars (FSRQs) and BL Lacertae objects (BL Lacs) is basically on the basis of rest-frame equivalent widths (EWs) of optical emission lines. One of the defining features of BL Lacs is their weak or absent emission lines \citep{UP95}. BL Lacs and FSRQs have EW $<5\rm{\AA}$ and EW $>5\rm{\AA}$, respectively, in the rest frame for optical emission lines, such as H$\beta\lambda 4861$, [O{\sc ii}]$\lambda 3727$, Mg{\sc ii}$\lambda 2798$, etc \citep[e.g.,][]{St91,UP95,Gh11,Sb12,GT15}.

Blazars show violent variability across the entire electromagnetic spectrum on different timescales from minutes to
years. Timescales and amplitudes of variations, and shapes of light curves (LCs), could shed light on some intrinsic properties of blazars, e.g., the sizes of emission regions, the masses of black holes, and the radiation mechanism \citep[e.g.,][]{Mi89,Xi02,LB15,Gu16,Fe17,Li17,Xi17,Yu17}. The variation timescales of blazars are divided into three classes. Intra-day variability (IDV) or micro-variability shows flux changes from minutes to less than one day \citep{WW95}. Short term variability shows variations from days to weeks, and long term variability (LTV) shows variations from
months to years \citep[e.g.,][]{Gu12,Li17}. Several models were proposed to explain the variability of blazars, such
as the shock-in-jet model \citep{MG85,Qi91} and the disk instability model \citep{Ka98}. The IDVs in blazars seem to
be important since the IDV timescales are likely related to the central supermassive black boles in blazars \citep{LB15,Fe17}. However, it is difficult to measure the masses of black holes in blazars due to the Doppler boosted emission from the jets of blazars.

S5 0716+714 is a typical BL Lac object discovered in 1979 \citep{Ku81}. \citet{WW95} found that the source is always in active states, and also similar results were reported in some works \citep[e.g.,][]{Ne02,Hu14}. Thus, S5 0716+714 may
be a good candidate for IDV researches. The optical IDV was studied extensively \citep[e.g.,][]{Gu09,Gu12,Da15,Ho17,Li17}. \citet{Ra10} reported a variation timescale as short as 15 minutes and \citet{Ma16} also obtained a variation timescale
of 17.6 minutes. Other variation timescales from tens of minutes to a few hours were reported in several works \citep[e.g.,][]{Gu09,Da15,Li17,Yu17}. The variability of 3C 273, the first quasar discovered in 1963 \citep{SH63}, was extensively investigated from radio to $\gamma$-rays \citep[e.g.,][]{Xi99,Fa09,Ab10a,Ab10c,Ka15,Xi17,Yu17}. 3C 273 was observed for more than 100 yr \citep[e.g.,][]{Vo13}, and optical variability on various timescales was reported. \citet{Fa09} reported the IDV timescales from 13 to 245 minutes for 3C 273. The variation timescales from 23.9 to 744
days were also found \citep{Fa14}. \citet{Da09} studied the spectrum variability, and the bluer-when-brighter behavior was obtained for the IDV and LTV. \citet{So08} suggested complicacy of the radiation mechanisms of the multiwavelength emission
for 3C 273. The photometry with high temporal resolution shorter than minutes may give more information for 3C 273 and
more constraints on its central supermassive black hole.

For S5 0716+714 and 3C 273, we carried out observations in the $I$ or $R$ bands from 2007 March 7 to 2012 May 16, and the observations were performed with high temporal resolution (30--60 s). Thus, we can investigate IDVs on shorter variation timescales in details. The structure of this paper is as follows. Section 2 gives observations and data reduction; Section
3 presents search for IDVs. Section 4 presents results, subsection 4.1 is for S5 0716+714, and subsection 4.2 is for 3C 273. Section 5 is for discussions and conclusion.

\section{OBSERVATIONS AND DATA REDUCTION}

The photometric observations of S5 0716+714 and 3C 273 were performed with the 1.02m optical telescope at Yunnan Observatories of Chinese Academy of Sciences from 2007 March 07 to 2012 May 16. Before 2009, the Princeton CCD chip
(1024 $\times$ 1024 pixels) of the 1.02m optical telescope covers a field of view (FOV) of $\sim$ 6.5 $\times$
6.5 arcmin$^2$, and the spatial scale is 0.38 aresec per pixel. For this CCD, the readout noise and gain are 3.9 electrons and 4.0 electrons/ADU, respectively. After 2009, the telescope was equipped with a new Andor DW436 CCD (2048 $\times$ 2048 pixels) camera at $f/13.3$ Cassergrain focus. The FOV of the CCD is $\sim$ 7.3 $\times$ 7.3 arcmin$^2$, and the projected angle on the sky of each pixel corresponds to 0.21 arcsec in both dimensions. The readout noise is 6.33 electrons, and the gain is 2.0 electrons/ADU. During the observations, standard Johnson-Couisns broadband filters were used \citep[e.g.,][]{Fe17}.

In order to improve the observation efficiency and to detect the optical variations with the shorter timescales, only one
band ($I$ or $R$) was observed in each night. 648 $I$ band CCD images of S5 0716+714 were obtained in six nights. For 3C 273, we observed 14 nights, and obtained 2305 CCD frames (611 in the $I$ band and 1694 in the $R$ band). Table 1 lists the complete observation log. The flat-field images were taken at twilight or dawn, and the bias frames were taken at the beginning and/or at the end of observations. Depending on the filters and weather conditions, the exposure times were set
as 30 s for S5 0716+714 and 30--80 s for 3C 273. All the CCD images were reduced by the standard IRAF procedures. For each night, the median of all the bias images was used to generate a master bias. Then, target images and flat-field images were subtracted by the master bias. After the bias correction, master flat-field images were generated by taking the median of all flat-field images in each band, and the target images were corrected by the master flat-field image. Before photometric reduction, we checked each image carefully. In the whole FOV, the background is nearly uniform, and the full width at half maximum (FWHM) is consistent for different stars. The values of FWHMs for most images are less than 2 arcsec. Thus, our bias and flat-field corrections are reliable. Aperture photometry was performed with the APPHOT task. S5 0716+714 and 3C 273 are point sources and our extraction aperture is determined by the FWHM. For each source, we chose 26 different aperture radii from 1.0 to 2.5 FWHM. Comparing the results of different aperture radii shows that the LCs are generally consistent with each other. The best signal-to-noise ratio (S/N) would be obtained with an aperture radius of 1.6 FWHM.

During our observations, several comparison stars are always located in the target FOV. For the most observable night (except S5 0716+714 on 2008 May 08), we can choose the same four comparison stars to calibrate the relevant target and characterize the uncertainties in the observations. Star2, star5, and star6 are always located in the FOV of 0716+714 on 2008 May 08, and are used in data reduction. Figure 1 shows the comparison stars star2, star3, star5, and star6 for S5 0716+714, and starC, starE, starG, and star1 for 3C 273. The magnitude calibration is performed as follows.

(1) There are several comparison stars that have been widely used in the previous works. Star2, star3, star5, and star6
for S5 0716+714 have been calibrated in \citet{Vi98}, and \citet{Sm85} has given the magnitude of starC, starE, and starG
for 3C 273. However, the transmittance of different filters might be slightly different, and the responses of different
CCDs are also different. Thus, we only adopt the brightness of the brightest star in the FOV to recalibrate other stars.
For each night, we measure the mean differential magnitude of every two comparison stars, $\overline{\Delta m_{\rm{i,j}}} (T_{\rm{m}})=\overline{star_{i}(T_{\rm{m}})- star_{j}(T_{\rm{m}})}$, where star$_{i}$ and star$_{j}$ are the instrumental magnitudes of the $i$th and $j$th comparison stars on the observation time series $T_{\rm{m}}$, respectively. We
find the mean value, $\overline{\Delta m_{\rm{i,j}}}$, of the same star pairs is nearly constant on the different nights for six pairs, i.e., $\mid \overline{\Delta m_{\rm{i,j}}}(T_{\rm{m}})- \overline{\Delta m_{\rm{i,j}}}(T_{\rm{n}})\mid\leq$
0.005 mag except for very few matching $\mid \overline{\Delta m_{\rm{i,j}}}(T_{\rm{m}})- \overline{\Delta m_{\rm{i,j}}} (T_{\rm{n}})\mid\leq$ 0.01 mag from 2007 to 2012. Therefore, the mean values are used to calibrate each comparison star.
We choose the brightness of star2 and starC as the standard flux of the image for S5 0716+714 and 3C 273, respectively.

(2) The brightness of comparison stars are considered to be constant, and the differential magnitude of any two stars
should be constant. Theoretically, we can use any star to calibrate the target. Nevertheless, the tracking accuracy
of the telescope, weather conditions, moon state, flat-field correction, and other unexpected reasons would influence the calibration of target. To avoid these effects, we calibrate the target by the different comparison stars ($Mag_{\rm{i}}=
BL - star_{\rm{i}} + std_{\rm{i}}$, $BL$ is the instrumental magnitude of target, $star_{\rm{i}}$ is the instrumental magnitude of the $i$th comparison star, and $std_{\rm{i}}$ is the calibrated magnitude of the $i$th comparison star). Then, we average any two calibrated results ($Mag_{\rm{ij}} = \overline{Mag_{\rm{i}} - Mag_{\rm{j}}}$), and shift the differential magnitude of the corresponding comparison stars to zero ($std_{\rm{ij}}= star_{\rm{i}}-star_{\rm{j}}- std_{\rm{i}}+ std_{\rm{j}}$). Depending on the variations of $std_{\rm{ij}}$, we exclude some preternatural data points of $Mag_{\rm{ij}}$. The threshold value is set as $\mid std_{\rm{ij}}\mid \leq$ 0.01 mag. Then, the left $Mag_{\rm{ij}}$ ($Mag$) are averaged as the final results. We also calculate the mean value of $std_{\rm{ij}}$ ($Std$), which can be used
to estimate the variability and systematic uncertainties of the target. Table 2 exhibits the results of sources and comparison stars.

The final errors of the target are calculated from two components. The first component is the Poisson errors,
$\sigma_{\rm{p}}$, of the target and comparison stars, and $\sigma_{\rm{p}}$ can be obtained from IRAF. Another component comes from some unexpected reasons mentioned in the previous paragraphs, and we attribute the relevant errors to the systematic uncertainties $\sigma_{\rm{s}}$, which can be given by $\sigma_{\rm{s}} = \mid Std \mid$. The final errors
are given by $\sigma = \sqrt{\sigma_{\rm{p}}^2+\sigma_{\rm{s}}^2}$ and are listed in Table 2.

\section{SEARCH FOR IDVs}
The variability amplitude ($Amp$) on a given night can be calculated by the definition of \citet{HW96}:
\begin{equation}
  Amp = 100 \times \sqrt{(Mag_{\rm{max}} - Mag_{\rm{min}})^{2} - 2\sigma^{2}} \%,
\end{equation}
where $Mag_{\rm{max}}$ and $Mag_{\rm{min}}$ are the maximum and minimum magnitudes within the LC, respectively, and
$\sigma$ can use the standard deviation of $Std$. Table 1 lists $Amp$ of the LCs, in which IDVs are detected. Two
standard statistical methods are used to investigate IDVs: the $F$-test and the one-way analysis of variance
\citep[ANOVA; e.g.,][]{de10,Ga12,Hu14,Ag15,Fe17}. If the LCs simultaneously satisfy the criteria of the $F$-test and
the one-way ANOVA test, the IDVs are tested further with a cross-correlation analysis.

The $F$-test have been widely used in detection of IDVs \citep[e.g.,][]{Hu14,Fe17,Xi17}. The value of $F$ is calculated
by comparing the variances of two samples and is defined as
\begin{equation}
  F = \frac{Var(Mag)}{Var(Std)},
\end{equation}
where $Var(Mag)$ is the variance of the calibrated magnitude of the blazar, and $Var(Std)$ is the variance of the calibrated comparison stars. The critical value of the $F$-test can be obtained by the $F$-statistic. The significance level is set at 0.01. Thus, if the $F$ value is larger than the critical value, the blazar is considered to be variable at the confidence level of 99\% (i.e., 2.6$\sigma$). Table 1 shows the $F$ values and the critical values. However, the $F$-test relies on the error of the target and comparison stars. Thus, another robust analysis method is necessary. The one-way ANOVA is a powerful tool to quantify the variability of blazars. \citet{de10} has investigated the one-way ANOVA in details and has shown that the one-way ANOVA is a powerful and robust method in the detection of IDVs. The one-way ANOVA does not depend on the error measurement but on the variability of blazars. The critical value of the one-way ANOVA test can be compared to the $F$-statistic \citep[see][for details]{de10,Fe17}. The one-way ANOVA tests are performed by grouping the data in sets of 20 individual observations \citep[see description in A.3 in][]{de10}. The method might be influenced by the intervals of the bins that are used to calculate ANOVA \citep[see A.3 in][]{de10}. So, we use five different bins of grouping 3, 4, 5, 6, and 7 data points for each night. If the data points in the last bin are less than those in the previous bin within the same LC, we merge them into the previous bin. As all the five groupings for the same LC detect the IDV, the LC is considered to have a IDV. The results of the one-way ANOVA and the critical values on the basis of grouping 7 data points are listed in Table 1. For comparison purposes, we also carry out the one-way ANOVA test on the curves of $Std$ of the comparison stars. The test results are listed in Table 1. As the one-way ANOVA test on $Std$ gives a variable result, the IDV of the target seems to be questionable (2007 March 07 and 2008 May 06 for S5 0716+714; 2008 May 08 for 3C 273). We will give further studies with cross-correlation analyses between the variations of the target and $Std$ for these three nights.

In order to avoid the illusive IDVs caused by the comparison stars, a discrete correlation function \citep[DCF; e.g.,][]{Ed88} is used to study correlations between the LCs of the target and the curves of $Std$. Correlation analyses
are used to test whether the variations of the target follow those of $Std$, i.e., the illusive variations of the target.
No correlations around zero time lags are expected for the relevant variations of the target and $Std$. If there are correlations around zero time lags, the target has the illusive variations. Correlation analyses are run for the relevant LCs when the target survives from the $F$-test and the one-way ANOVA test (see Table 1). The LCs in Figures 3 and 4 are non-uniformly sampled. The $z$-transformed discrete correlation function \citep[ZDCF;][]{Al97} is a binning type of method as an improvement of the DCF technique, but it has a notable feature in that the data are binned by equal population rather than equal bin width as in the DCF \citep[e.g.,][]{LB11}. The ZDCF is more robust than the DCF when applied to unequally sampled LCs \citep[see][]{LB11}. The ZDCF results are presented in Figures 5 and 6. For S5 0716+716, there is no correlation on 2007 March 07, but there is a correlation on 2008 May 06. There is no correlation on 2008 May 08 for 3C 273. Thus, the IDVs of these three nights are questionable even if these LCs of the two targets survive the $F$-test and the one-way ANOVA test. So, these three nights are not considered to have reliable IDVs. Moreover, there is a correlation on 2008 May 06 for 3C 273 (see Figure 6). Finally, four and five reliable IDVs survive the ZDCF cross-correlation zero lag test for S5 0716+714 and 3C 273, respectively.

\section{RESULTS}

The long-term LCs are displayed in Figure 2. Figures 3 and 4 show the LCs that survive the $F$-test and the one-way ANOVA test for S5 0716+714 and 3C 273, respectively. The details of the IDV LCs are as follows below.

\subsection{S5 0716+714}
The LC on 2008 May 06 cannot survive the ZDCF cross-correlation zero lag test for S5 0716+714. The $R$-band magnitudes are converted to linear fluxes of $F$ using the formula $F=3.08\times 10^{-0.4\times R+3}$ $\rm{Jy}$, and the $I$-band magnitudes are converted to linear fluxes of $F$ using the formula $F=2.55\times 10^{-0.4\times I+3}$ $\rm{Jy}$ \citep[e.g.,][]{Fe18}. During our observations, S5 0716+714 was active, and the IDVs were detected in 5 out of 6 days. Rising and declining phases were observed on 2007 March 08 and 09, respectively (see Figure 3). On 2007 March 08, it was almost monotonically increasing by $\Delta I \approx$ 0.05$\pm$0.01 mag in $\approx 0.09$ days. On the following day, S5 0716+714 faded by $\Delta I \approx$ 0.05$\pm$0.01 mag in $\approx$ 0.05 days. The flare of S5 0716+714 on 2008 May 08 can be fitted by a third-order polynomial with a reduced Chi-square $\chi^2_{\nu}=0.790$ (see Figure 7a).

On 2008 May 07, S5 0716+714 darkens by $\Delta I\approx$ 0.06$\pm$0.01 mag in $\sim$ 0.05 days. On 2008 May 08, we detected successive rising, declining, and rising variations with magnitude changes of $\Delta I\approx$ 0.08$\pm$0.01 mag (see Figure 3). First, S5 0716+714 brightens slowly by $\Delta I\approx$ 0.08$\pm$0.01 mag in 53.4 minutes and darkens fast by $\Delta I\approx$ 0.07$\pm$0.01 mag in 13.2 minutes. Second, a little flare varies by $\Delta I\approx$ 0.04$\pm$0.01 mag in 13.6 minutes. Finally, S5 0716+714 brightens fast by $\Delta I\approx$ 0.08$\pm$0.01 mag in 6.6 minutes and darkens by $\Delta I\sim$ 0.04$\pm$0.01 mag in 5.4 minutes. From MJD = 595.06049 to 595.03814 (MJD = JD-2454000), S5 0716+714 brightens by $\Delta I\approx$ 0.05$\pm$0.01 mag in 19.0 minutes and darkens by $\Delta I\approx$ 0.07$\pm$0.01 mag in 13.2 minutes. This variation has a duration 32.2 minutes that can give the minimum timescale of variations during our observations of S5 0716+714. $Amp$ on 2008 May 07 and 08 are 10.8\% and 9.8\%, respectively. The long term variation amplitude of S5 0716+714 is 0.75 $\pm$ 0.01 mag in the $I$ band (see Figure 2). However, the poor sampling and the single color limit us to investigating the LTV.

\subsection{3C 273}
3C 273 was observed in the $I$ or $R$ bands on 14 nights, and only five nights survive three tests of IDVs. Though the LC
on 2008 May 06 survives the $F$-test and the one-way ANOVA test, it cannot survive the ZDCF cross-correlation zero lag test. Only one night is found to be variable in the $I$ band, and $\Delta I \approx$ 0.04$\pm$0.01 mag on 2008 May 07. Other 4
IDV events are detected in the $R$ band. On 2010 May 18, $Amp$ of 3C 273 is 3.3\% in the $R$ band. At the beginning of the LC on 2011 May 07, the source quickly brightens by $\Delta R \approx$ 0.05$\pm$0.01 mag in 0.0185 days (26.6 minutes) and then is almost at a constant level (see Figure 4). For the night of 2011 May 09, it darkens from MJD = 1690.68818 to 1690.70038 and brightens from MJD = 1690.70038 to 1690.70363. This dip shows $\Delta R \approx$ 0.04$\pm$0.01 mag in 22.2 minutes. The next dip darkens from MJD = 1690.70363 to 1690.71014, and brightens from MJD = 1690.71014 to 1690.71095. This dip shows $\Delta R \approx$ 0.05$\pm$0.01 mag in 10.5 minutes. The next three dips are sharper. The third dip darkens from MJD = 1690.81278 to 1690.81453 and brightens from MJD = 1690.81453 to 1690.81904. This dip varies by $\Delta R \approx$ 0.05$\pm$0.01 mag in 9.0 minutes. The fourth dip varies by $\Delta R \approx 0.05\pm 0.01$ mag from MJD = 1690.82719 to 1690.83009 and lasts for 4.2 minutes. The fifth dip varies by $\Delta R \approx 0.05\pm 0.01$ mag from MJD = 1690.88119
to 1690.88608 and lasts for 7.0 minutes.

The LC on 2011 May 10 has a variation amplitude of $\Delta R \approx$ 0.06$\pm$0.01 mag. After MJD = 1691.81, there is a flare with $\Delta R \approx$ 0.05$\pm$0.01 mag (see Figures 4 and 7b). The flare has a basically complete profile, consists of 29 data points, and lasts for 0.40 hr (see Figure 7b). The flare duration can give the minimum timescale of variations during our observations of 3C 273. After this flare, there are seven data points in a darkening phase around MJD = 1691.86, and these points can be fitted linearly with a Pearson's correlation coefficient $r=0.963$ at the confidence level of 99.95\%. (see Figure 7b). This darkening phenomenon in 3C 273 is similar to that in Mrk 501 \citep[see Figure 6 in][]{Fe17}. On 2011 May 10, the flare of 3C 273 can be fitted by a third-order polynomial with $\chi^2_{\nu}=0.504$ (see Figure 7b). For a relatively complete flare, the variation timescale could be estimated by the interval between the local minima at the adjacent valleys in the LC \citep[see][for details]{Fe17}. This variation timescale is basically consistent with the duration of the flare.

For the long term LCs of 3C 273 in the $I$ band, a rising phase from 2007 March to 2008 May has $\Delta I\approx$ 0.36$\pm$0.01 mag, and a declining phase from 2008 May to 2009 May gives $\Delta I \approx$ 0.54$\pm$0.01 mag. For the
long term LCs of 3C 273 from 2009 May to 2012 May, our observational data basically follow the variation trend of the LC from SMARTS\footnote{http://www.astro.yale.edu/smarts/fermi} in the $R$ band \citep{Bo12} (see Figure 8). The IDV behaviors are found on 2011 May 07, 09, and 10, when 3C 273 is nearly at the brightest of the rising phase from 2010 May to 2011 May. 3C 273 darkens from 2008 May to 2009 May in our observations. A similar darkening trend from MJD $\sim$ 500 to 900 appears in the LCs from SMARTS \citep{Bo12}.

\section{DISCUSSION AND CONCLUSIONS}

We monitored BL Lac object S5 0716+714 and FSRQ 3C 273 with high time resolutions (30--60 s) from 2007 March 07 to 2012
May 16, and IDV behaviors are found in the $I$ or $R$ bands for these two sources. The minimum timescales are 0.54 and
0.40 hr for S5 0716+714 and 3C 273, respectively. \citet{Yu17} reported that the minimum timescales of S5 0716+714
and 3C 273 are 0.29 and 0.59 hr, respectively. Our results are consistent with theirs in the order of magnitude. These variability timescales could give upper limits to the sizes of emission regions, $D \lesssim c \Delta t_{\rm{min}}^{\rm{ob}} \delta /(1+z)$, where $c$ is the speed of light, $\Delta t_{\rm{min}}^{\rm{ob}}$ is the observed minimum timescale of variability, $\delta$ is the Doppler factor, and $z$ is the redshift of the source. S5 0716+714 is at $z=0.31$ \citep{Ni08,Da13}, and its $\delta \sim$ 10.8 \citep{Sa10}. Thus, $D \lesssim 4.78\times10^{14}$ cm for S5 0716+714. The values of $z$ and $\delta$ are 0.158 and 16.8 for 3C 273 \citep{Sa10}, respectively, and $D\lesssim 6.27\times 10^{14}$ cm.

The close correlations between the flares of different bands indicate that the IDV is an intrinsic phenomenon \citep{WW95}. Some models were proposed to study the underlying connections between the timescales of variations and the masses of black holes, $M_{\bullet}$ \citep[e.g.,][]{Ab82,Mi89,Xi02,LB15}. The observed minimum timescales, $\Delta t_{\rm{min}}^{\rm{ob}}$, of variability were generally used to estimate $M_{\bullet}$ for AGNs \citep[e.g.,][]{Ab82,Mi89,Xi02,Xi05,Da15,LB15}. Models based on accretion disk were proposed to connect $\Delta t_{\rm{min}}^{\rm{ob}}$ and $M_{\bullet}$ for non-blazar-like AGNs \citep[e.g.,][]{Ab82,Mi89,Xi02}. \citet{LB15} proposed a new sophisticated model based on a blob in a relativistic jet to limit $M_{\bullet}$ for blazars, and the upper limits to $M_{\bullet}$ are given by
\begin{mathletters}
  \begin{eqnarray}
    M_{\bullet}\lesssim 5.09\times 10^4 \frac{\delta \Delta t_{\rm{min}}^{\rm{ob}}}{1+z}M_{\odot} \/\ \/\ (\/\ j\sim 1),\\
    M_{\bullet}\lesssim 1.70 \times 10^4 \frac{\delta \Delta t_{\rm{min}}^{\rm{ob}}}{1+z}M_{\odot} \/\ \/\ (\/\ j=0),
  \end{eqnarray}
\end{mathletters}
where $\Delta t_{\rm{min}}^{\rm{ob}}$ is in units of seconds, $j = J/J_{\rm{max}}$ is the dimensionless spin parameter
of a black hole with the maximum possible angular momentum of $J_{\rm{max}} = G M_{\rm{\bullet}}^2/c$, and $G$ is the gravitational constant. Equations (3a) and (3b) can be applied to the Kerr and Schwarzchild black holes, respectively. For S5 0716+714, we have $M_{\bullet}\lesssim 10^{8.43} M_{\odot}$ for the Schwarzchild black hole and $M_{\bullet}\lesssim 10^{8.91} M_{\odot}$ for the Kerr black hole. \citet{LL03} used the optical luminosity to get a mass of $M_{\bullet} = 10^{8.10} M_{\odot}$, which is consistent with our results. For 3C 273, we have $M_{\bullet} \lesssim 10^{9.02} M_{\odot}$ derived with equation (3a). \citet{Ka00} obtained $M_{\bullet} = 0.235^{+0.037}_{-0.033}$--$0.550^{+0.089}_{-0.079}\times 10^{9} M_{\odot}$ from the reverberation mapping of the Balmer lines, which are consistent with our result. \citet{PT05} obtained $M_{\bullet} = 2.44^{+0.51}_{-0.30}\times 10^{9} M_{\odot}$ from the reverberation mapping of the Balmer lines
and the Ly$\alpha$ and C{\sc iv} lines, and generally, this mass is larger than other measurements in the literatures. Also, this mass is larger than our result. \citet{Pe04} also obtained a reverberation-based mass of $M_{\bullet} = (8.86\pm1.87) \times 10^{8} M_{\odot}$, which is consistent with the upper limit of $M_{\bullet} \lesssim 10^{9.02} M_{\odot}$. \citet{Zh18} derived $M_{\bullet}=10^{9.0\pm0.8}M_{\odot}$ from the correlation between the host bulge stellar mass and the black hole mass and obtained a reverberation-mapped mass of $M_{\bullet}= 4.1^{+0.3}_{-0.4}\times 10^{8}M_{\odot}$. \citet{St18} inferred a mass of $M_{\bullet}=(2.6\pm1.1)\times 10^{8}M_{\odot}$ from GRAVITY interferometry observation data of the Paschen-$\alpha$ line for 3C 273. These new measurements are consistent with our result of $M_{\bullet} \lesssim 10^{9.02} M_{\odot}$. Thus, our estimates of $M_{\bullet}$ for S5 0716+714 and 3C 273 are consistent with a model in which their optical IDVs are generated from jets.

Except for the jet origin of optical IDVs, an alternative way can explain optical IDVs, e.g., accretion disks \citep[e.g.,][]{Ag16}. Though the accretion disk instability can explain some phenomena in the optical--X-ray bands,
it cannot explain the radio IDV behaviors \citep[e.g.,][]{WW95}. Thermal emission from the accretion disk is not found in multiwavelength SEDs of S5 0716+714 \citep[e.g.,][]{Li14}. The optical emission of S5 0716+714 is from the synchrotron process of relativistic electrons in relativistic jets, and the $\gamma$ rays are interpreted as the inverse Compton (IC) scattering of soft photons by the relativistic electrons that produce the optical emission \citep[e.g.,][]{Li14}. Thus,
the ionizing radiation is so weak that broad emission lines are not observable, even though broad emission line
region exists in S5 0716+714. Then, its optical spectra will be featureless. Broad emission lines were observed only in a few BL Lac objects \citep[e.g.,][]{Ce97,Ca99}. Accretion rates are very low for BL Lac objects \citep[e.g.,][]{Ca02,Xu09}. The absence of broad emission lines in most of BL Lac objects may be due to the very weak emission of the accretion disk. \citet{Ni08} used the host galaxy of S5 0716+714 as the "standard candle" to derive its redshift of $z = 0.31 \pm0.08$ during its low state. BL Lac object PKS 0537-441 shows an interesting event in the $J$ band with a duration of $\sim$ 25 minutes \citep{Im11}. In both the low and high states, its emission appears to be dominated by a jet, and no evidence of a thermal emission is apparent. Its SEDs are produced by the synchrotron and IC processes within a jet \citep{Pi07}. For TeV $\gamma$-ray BL Lac object Mrk 501, the optical emission is neither the thermal component from accretion disk nor the nonthermal component from a jet \citep{Ah17}. The optical emission is dominated by the host galaxy, and the UV emission
is from the jet for Mrk 501. Thus, it is not possible that the optical IDV behaviors are from accretion disk for BL Lac objects with the featureless optical spectra.

The featureless optical spectrum is the typical characteristic of BL Lac objects. On the contrary, quasars show many
strong broad emission lines. 3C 273 has strong broad emission lines of the Balmer series and Ly$\alpha$. The broadband
SED of 3C 273 shows a prominent blue-bump around the UV--optical regime \citep{Tu99}. The blue-bump may be attributed to 
the Fe{\sc ii} and Balmer line and continuum emission \citep{Pa98}. If the blue-bump is the thermal emission from the accretion disk, Equations 3(a) and (b) are not appropriate to estimate $M_{\bullet}$ for 3C 273. For the Kerr black hole, \citet{Xi02} deduced a formula for the accretion disk from \citet{Ab82}:
\begin{equation}
 M_{\bullet}\lesssim 1.62\times 10^4 \frac{ \Delta t_{\rm{min}}^{\rm{ob}}}{1+z}M_{\odot}.
\end{equation}
The flare duration of 0.40 hr and equation (4) give $M_{\bullet} \lesssim 10^{7.30} M_{\odot}$ for 3C 273. This upper limit of $M_{\bullet}$ is much lower than masses of $M_{\bullet}=10^{8.37}$--$10^{9.39} M_{\odot}$ obtained in the literatures. It may be not possible that the blue-bump is the thermal emission from the accretion disk for 3C 273. Thus, the flare with a duration of 0.40 hr is likely produced from the relativistic jets in 3C 273. Equation 3(a) gives a reasonable constraint on $M_{\bullet}$ for 3C 273. The shock-in-jet model, the most frequently used model to explain the IDV behaviors that may be directly related to shock processes in a jet, is based on a relativistic shock propagating down a jet and interacting with a highly nonuniform portion in the jet flow \citep[e.g.,][and references therein]{Na12,Su12,Ma14,Sa15}. As the relativistic shock passes through a blob in the jet, an IDV behavior may be produced.

In summary, the photometric observations with high temporal resolution in the $I$ or $R$ bands are used to search for
the optical IDV behaviors of S5 0716+714 and 3C 273. The observations were performed with the 1.02 m optical telescope
from 2007 March 07 to 2012 May 16. We obtained 687 $I$ band CCD images in six nights for S5 0716+714. For 3C 273, we obtained 2283 CCD frames (622 frames in the $I$ band and 1661 frames in the $R$ band) in 14 nights. The one-way ANOVA test is carried out on $Std$ of the comparison stars. There are IDVs of $Std$ for 3 out of the 20 nights. The IDVs of the target are not reliable if the one-way ANOVA test gives IDVs for the target and $Std$. Finally, four and five reliable IDVs survive the $F$-test, the one-way ANOVA test, and the ZDCF cross-correlation zero lag test for S5 0716+714 and 3C 273, respectively. Optical IDVs with flare durations of 0.54 and 0.40 hr are found for S5 0716+714 and 3C 273, respectively. Based on equation (3a) and $\Delta t_{\rm{min}}^{\rm{ob}}$ taken as flare durations, we estimate upper bounds to $M_{\bullet}$. $M_{\bullet}\la 10^{8.91} M_{\odot}$ and $M_{\bullet} \la 10^{9.02} M_{\odot}$ are given for S5 0716+714 and 3C 273, respectively. Our mass measurements are consistent with most of the measurements reported in the literatures, except for $M_{\bullet} = 2.44^{+0.51}_{-0.30}\times 10^{9} M_{\odot}$ for 3C 273 \citep{PT05}, which is generally larger than other measurements. This supports that these optical IDVs are from the jets, rather than the accretion disks, of S5 0716+714 and 3C 273. In addition, sharp dips are found in the LC on 2011 May 9 for 3C 273, and show $\Delta R \thickapprox$ 0.05$\pm$0.01 mag.

\acknowledgements {We are grateful to the anonymous referee for constructive comments leading to significant improvement
of this paper. Thanks for the helpful comments from the ApJ statistics editor. H.T.L. and H.C.F. thank the helpful discussions of Prof. Ji-Rong Mao and Dr. Zhi-Xiang Zhang. We thank the financial support of the Key Research Program
of the CAS (grant No. KJZD-EW-M06), the National Natural Science Foundation of China (NSFC; grant Nos. 11433004 and 11573067), and the Ministry of Science and Technology of China (2016YFA0400700). We also thank the financial support of 
the NSFC (grant No. 11273052) and the CAS Interdisciplinary Innovation Team.}

\clearpage

\begin{deluxetable}{lccccccccc}
  \tablecolumns{13}
  \setlength{\tabcolsep}{3pt}
  \tablewidth{0pc}
  \tablecaption{Observation log and results of IDV observations of blazars}
  \tabletypesize{\scriptsize}
  \tablehead{
  \colhead{Date}                                             &
  \colhead{Target}                                           &
  \colhead{Filter}                                          &
  \colhead{$N$}                                                &
  \colhead{$F$}                                              &
  \colhead{$F$(99)}                                          &
  \colhead{ANOVA}                                            &
  \colhead{ANOVA(99)}                                        &
  \colhead{Variable}                                         &
  \colhead{$Amp$(\%)}                                         \\

\colhead{(1)}&\colhead{(2)}&\colhead{(3)}&\colhead{(4)}&\colhead{(5)}&\colhead{(6)}&
\colhead{(7)}&\colhead{(8)}&\colhead{(9)}&\colhead{(10)}}
\startdata
2007-03-07&	S5 0716+714	& $I$ & 105 & 9.85  & 1.58 & 7.25  & 2.29 & Yes &	4.0	 \\
          & $Std^{\ast}$ &   &     &       &      & 2.62 & 2.29 &  Yes  &     \\
2007-03-08&	S5 0716+714	& $I$ & 127 &	28.04 &	1.52 & 12.43 & 2.14 & Yes &	7.2	 \\
          & $Std^{\ast}$ &   &     &       &       & 1.98 & 2.14 &  No    &     \\
          &	3C 273	    & $I$ & 85  & 4.22 &	1.67 & 1.87 & 2.50	& No  &	...	 \\
          &$Std^{\ast}$ &   &     &       &      & 1.51  & 2.50 &  No  &     \\
2007-03-09&	S5 0716+714	& $I$ & 152 &	18.57 &	1.46 & 19.19 & 2.02 & Yes &	5.9	 \\
         & $Std^{\ast}$ &   &     &       &      & 1.03  & 2.02 & No  &     \\
          &	3C 273	    & $I$ & 87 &	1.52  &	1.66 & 1.35	 & 2.49	& No  &	...	 \\
         &$Std^{\ast}$ &    &     &       &      & 0.73  & 2.49 & No  &     \\
2008-05-06&	S5 0716+714	& $I$ & 90 &	13.58 &	1.64 & 12.73 & 2.48	& Yes &	4.9	 \\
         & $Std^{\ast}$ &   &     &       &      & 3.71   & 2.48 & Yes &     \\
          &	3C 273	    & $I$ & 95  &	4.52  &	1.62 & 2.43  & 2.41	& Yes &	3.6  \\
         & $Std^{\ast}$ &   &     &       &       & 2.32  & 2.41 & No  &     \\
2008-05-07&	S5 0716+714	& $I$ & 96 &	42.28 &	1.62 & 21.6 & 2.41 & Yes &	10.8 \\
         & $Std^{\ast}$ &   &     &       &      & 0.93  & 2.41 &  No  &     \\
          &	3C 273	    & $I$ & 94  &	4.27  &	1.63 & 2.49 & 2.41 & Yes &	4.0	 \\
         & $Std^{\ast}$ &   &     &       &       & 2.16 & 2.41 & No  &     \\
2008-05-08&	S5 0716+714	& $I$ & 78 &	22.01 &	1.71 & 6.01 & 2.60 & Yes &	9.8  \\
         & $Std^{\ast}$ &   &     &       &      & 0.61  & 2.60 & No &     \\
          &	3C 273	    & $I$ & 130 &	3.27  &	1.51 & 2.42  & 2.13 & Yes &	4.8  \\
         &$Std^{\ast}$ &   &     &       &       & 2.56  & 2.13 & Yes &     \\
2009-05-16&	3C 273	    & $I$ & 120 &	8.65  &	1.54 & 0.41  & 2.18	& No  &	...	 \\
         & $Std^{\ast}$ &   &     &       &      & 1.47  &  2.18 & No &     \\
2010-05-15&	3C 273	    & $R$ & 143 &	1.62  &	1.48 & 0.92  & 2.06 & No	  &	...	 \\
         & $Std^{\ast}$ &   &     &       &       & 5.96  & 2.06 & Yes &     \\
2010-05-16&	3C 273	    & $R$ & 92  &	3.77  &	1.63 & 1.35  & 2.42 & No	  &	...	 \\
         &$Std^{\ast}$ &   &     &       &      & 1.45  & 2.42 & No &     \\
2010-05-17&	3C 273	    & $R$ & 185 &	1.73  &	1.41 & 1.81  & 1.89 & No	  &	...	 \\
         & $Std^{\ast}$ &   &     &       &      & 0.47  & 1.89 &  No &     \\
2010-05-18&	3C 273	    & $R$ & 302 &	4.53  &	1.31 & 4.50  & 1.66 & Yes &	3.3	 \\
         & $Std^{\ast}$ &   &     &       &      & 1.10  & 1.66 &   No &     \\
2011-05-07&	3C 273	    & $R$ & 225 &	4.71  &	1.37 & 6.25  & 1.79 & Yes &	5.2	 \\
         &$Std^{\ast}$  &   &     &       &      &  2.27  & 1.79 & Yes &     \\
2011-05-09&	3C 273	    & $R$ & 307 &	18.29 & 1.31 & 2.61  & 1.65 & Yes &	11.7 \\
         & $Std^{\ast}$ &    &     &       &      & 0.66 & 1.65 & No  &     \\
2011-05-10&	3C 273	    & $R$ & 258 &	28.45 &	1.34 & 5.54  & 1.73 & Yes &	13.1 \\	
         &$Std^{\ast}$ &    &     &       &      & 1.55 & 1.73 & No &     \\
2012-05-16&	3C 273	    & $R$ & 182 &	7.37  &	1.42 & 1.12  & 1.90	& No  &	...	 \\
         & $Std^{\ast}$ &   &     &       &      &  1.44  & 1.90 & No  &     \\

\enddata
\tablecomments{Column 1: date of observation; Column 2: target, and $Std^{\ast}$ denotes the vertically moved $Std$ in Figures 2, 3, 4, and 6; Column 3: filter used in observations; Column 4: number of observations of each night; Column 5:
$F$ of the $F$ test for the observation data; Column 6: $F$(99) is the critical $F$ value at a 99\% confidence level;
Column 7: ANOVA of the ANOVA test for the observation data; Column 8: ANOVA(99) is the critical ANOVA value at a 99\% confidence level; Column 9: label of IDV; Colume 10: $Amp$ of IDV.}
\end{deluxetable}


\begin{deluxetable}{cccccccccc}
  \tablecolumns{7}
  \setlength{\tabcolsep}{3pt}
  \tablewidth{0pc}
  \tablecaption{Observational data for blazars}
  \tabletypesize{\scriptsize}
  \tablehead{
  \multicolumn{3}{c}{S5 0716+714}           &
  \multicolumn{1}{c}{}                      &
  \multicolumn{6}{c}{3C 273}                 \\
  \cline{1-3} \cline{5-10}
  \multicolumn{3}{c}{$I$}                     &
  \multicolumn{1}{c}{}                      &
  \multicolumn{3}{c}{$I$}                     &
  \multicolumn{3}{c}{$R$}                     \\
  \cline{1-3} \cline{5-10}

  \colhead{JD - 2454000}                             &
  \colhead{Mag}                                     &
  \colhead{$Std$}                                   &

  \colhead{}                                        &

  \colhead{JD - 2454000}                             &
  \colhead{Mag}                                     &
  \colhead{$Std$}                                    &

  \colhead{JD - 2454000}                             &
  \colhead{Mag}                                      &
  \colhead{$Std$}
}
\startdata
166.991528 &12.675 $\pm$ 0.016 &-0.005& &168.270185 & 12.058 $\pm$ 0.007 &-0.003& 1331.736759 &12.675 $\pm$ 0.006&-0.000 \\
166.993299 &12.675 $\pm$ 0.013 &-0.006& &168.271377 & 12.056 $\pm$ 0.005 &-0.000& 1331.738762 &12.666 $\pm$ 0.003& 0.001 \\
166.993958 &12.669 $\pm$ 0.009 &-0.000& &168.272060 & 12.057 $\pm$ 0.007 &-0.005& 1331.740336 &12.672 $\pm$ 0.007& -0.007 \\
166.994630 &12.663 $\pm$ 0.008 &-0.002& &168.272743 & 12.057 $\pm$ 0.006 &-0.004& 1331.741377 &12.673 $\pm$ 0.005& -0.004 \\
166.995278 &12.662 $\pm$ 0.009 &-0.002& &168.273553 & 12.067 $\pm$ 0.006 &0.001& 1331.742419  &12.670 $\pm$ 0.006& -0.005 \\
166.995972 &12.675 $\pm$ 0.010 &-0.004& &168.274259 & 12.069 $\pm$ 0.006 &0.003& 1331.743461  &12.680 $\pm$ 0.005&-0.003 \\
...        &...                & ...  & &	...     &  ...	             &...  & ...          & ... & ...  \\

\enddata
\tablecomments{This table is available in its entirety in a machine-readable form in the online journal. A portion is
shown here for guidance regarding its form and content. Mag denotes magnitude and corresponding error. $Std$ denotes the mean value of $std_{\rm{ij}}$ of the comparison stars -- see the text. The curves denoted by triangles in Figures 2, 3, 4,
and 6 correspond to the vertically moved $Std$. The uncertainty of each point is $\sigma = \sqrt{\sigma_{\rm{p}}^2+ \sigma_{\rm{s}}^2}$ -- see the text.}
\end{deluxetable}


\begin{figure*}
 \begin{center}
  \includegraphics[angle=0,scale=0.35]{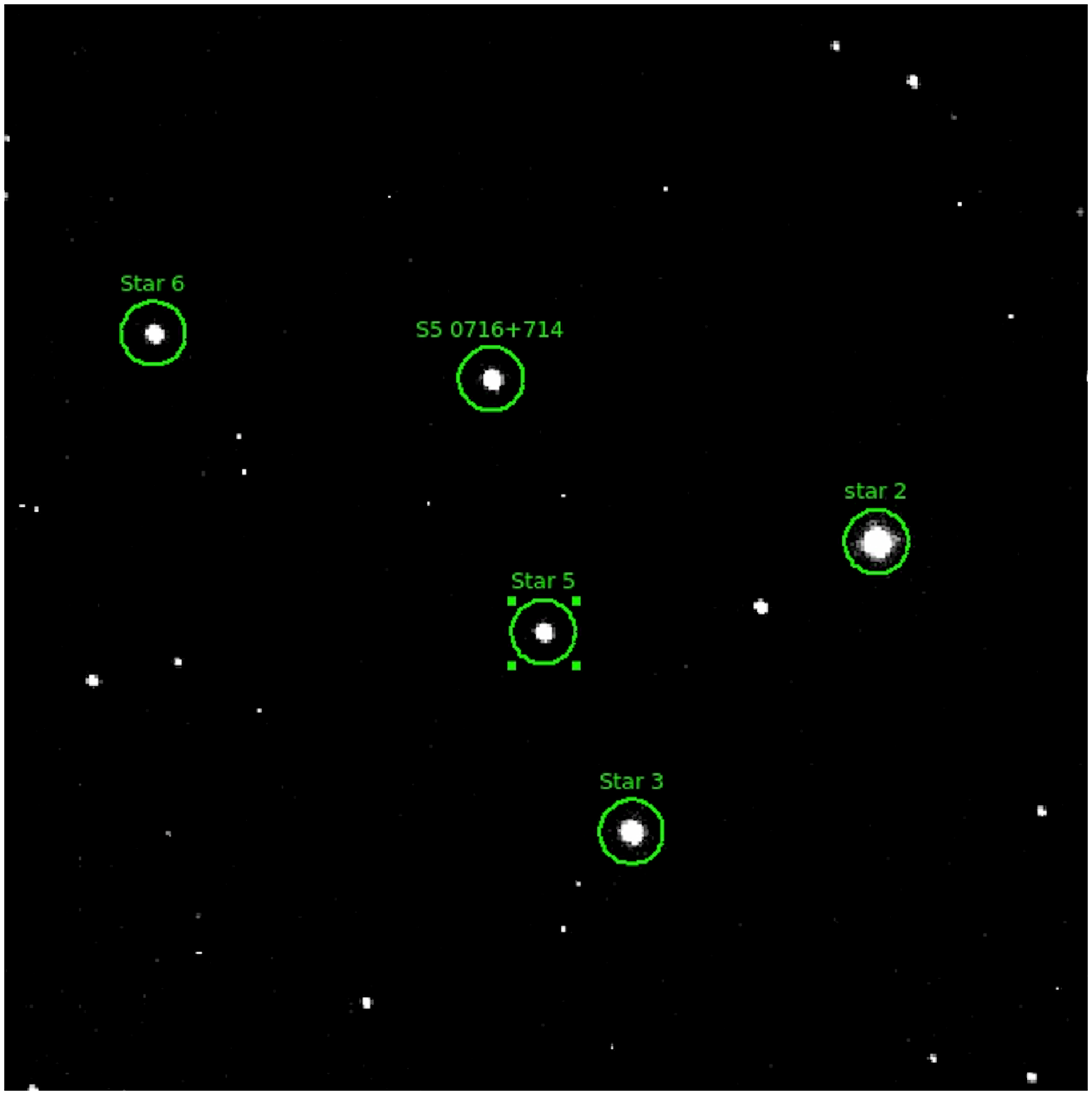}
  \includegraphics[angle=0,scale=0.35]{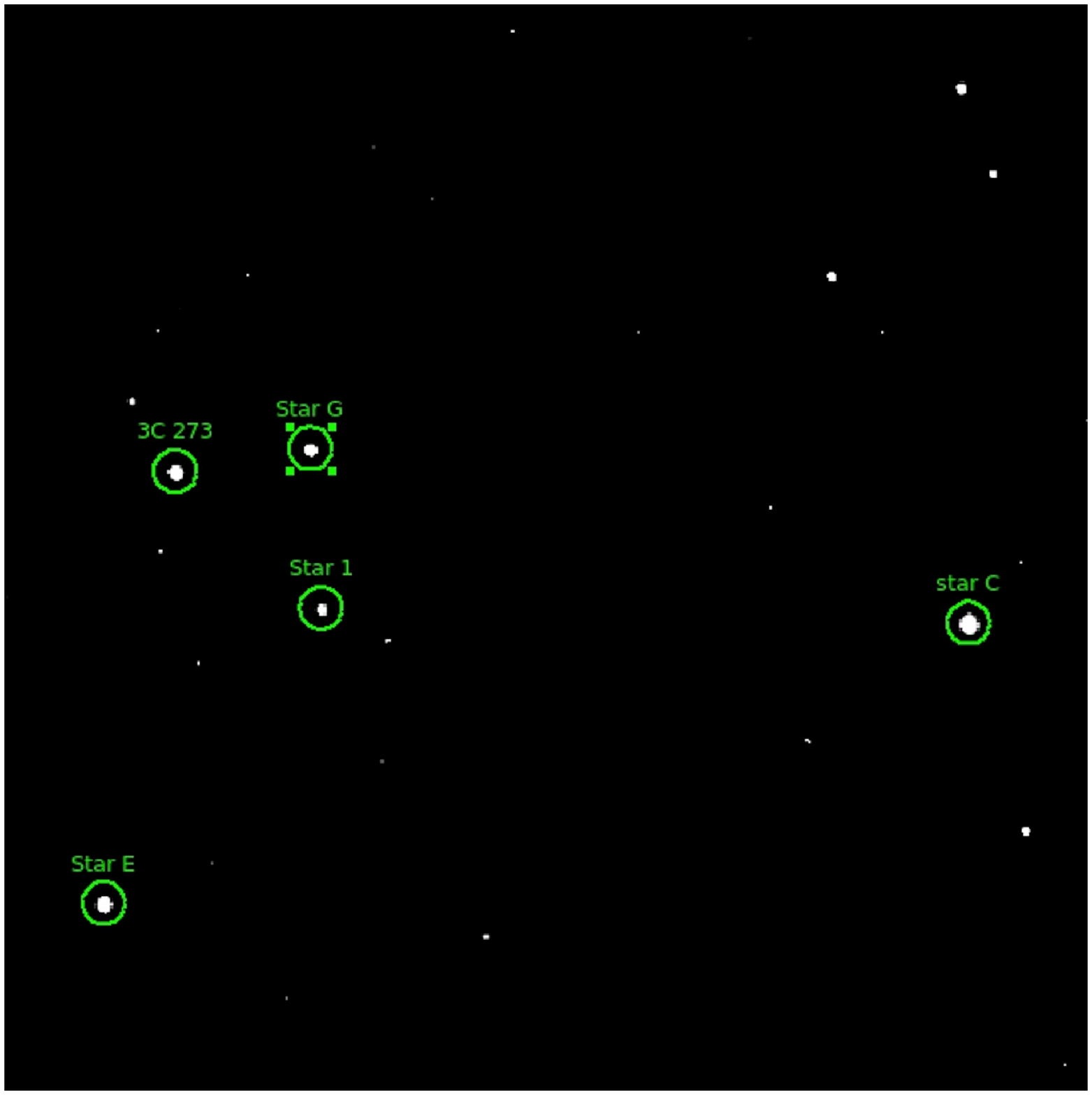}
 \end{center}
 \caption{Individual images of S5 0716+714 (left) and 3C 273 (right).}
  \label{fig1}
\end{figure*}


\begin{figure*}
 \begin{center}
  \includegraphics[angle=-90,scale=0.3]{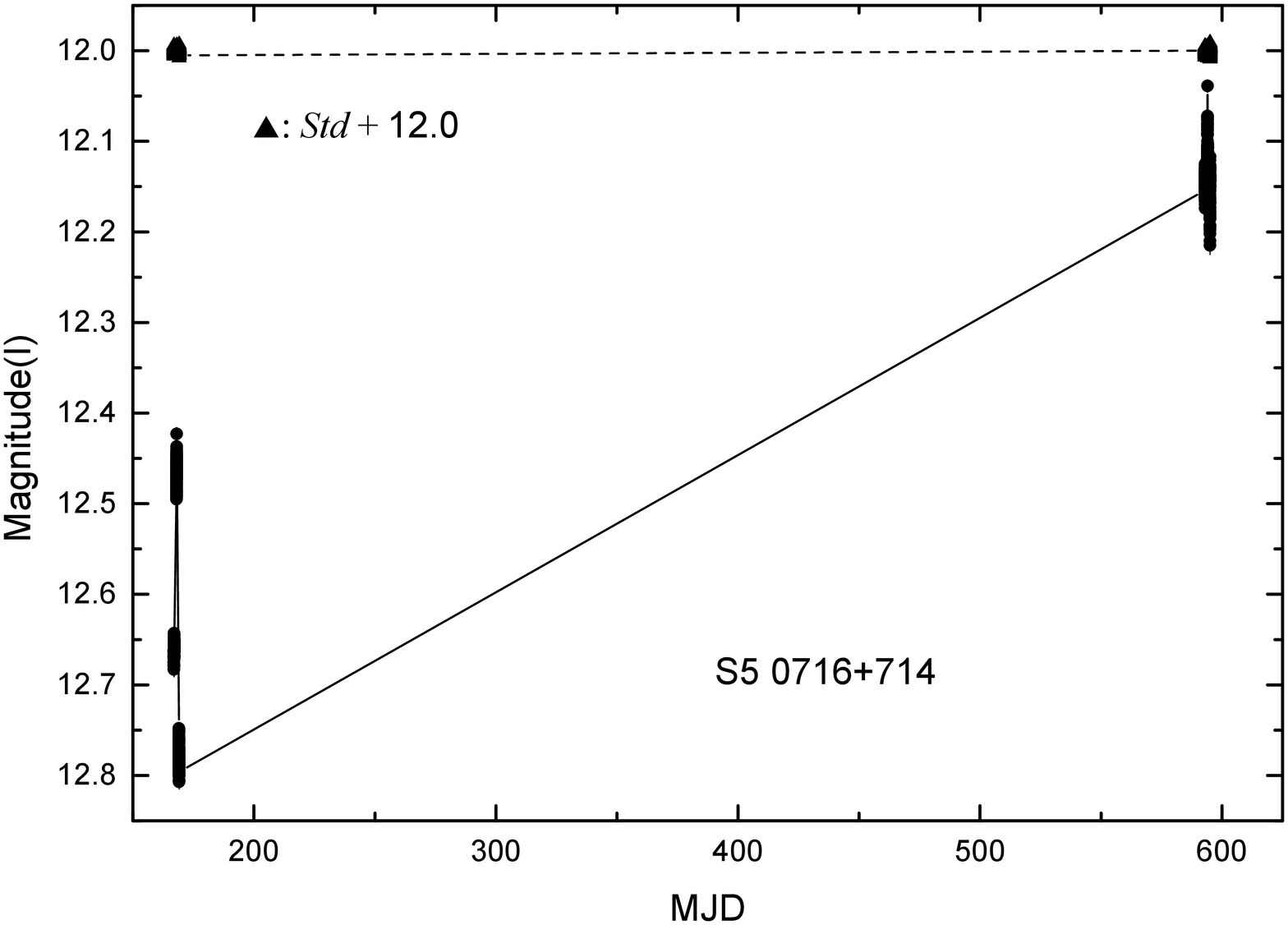}
  \includegraphics[angle=-90,scale=0.3]{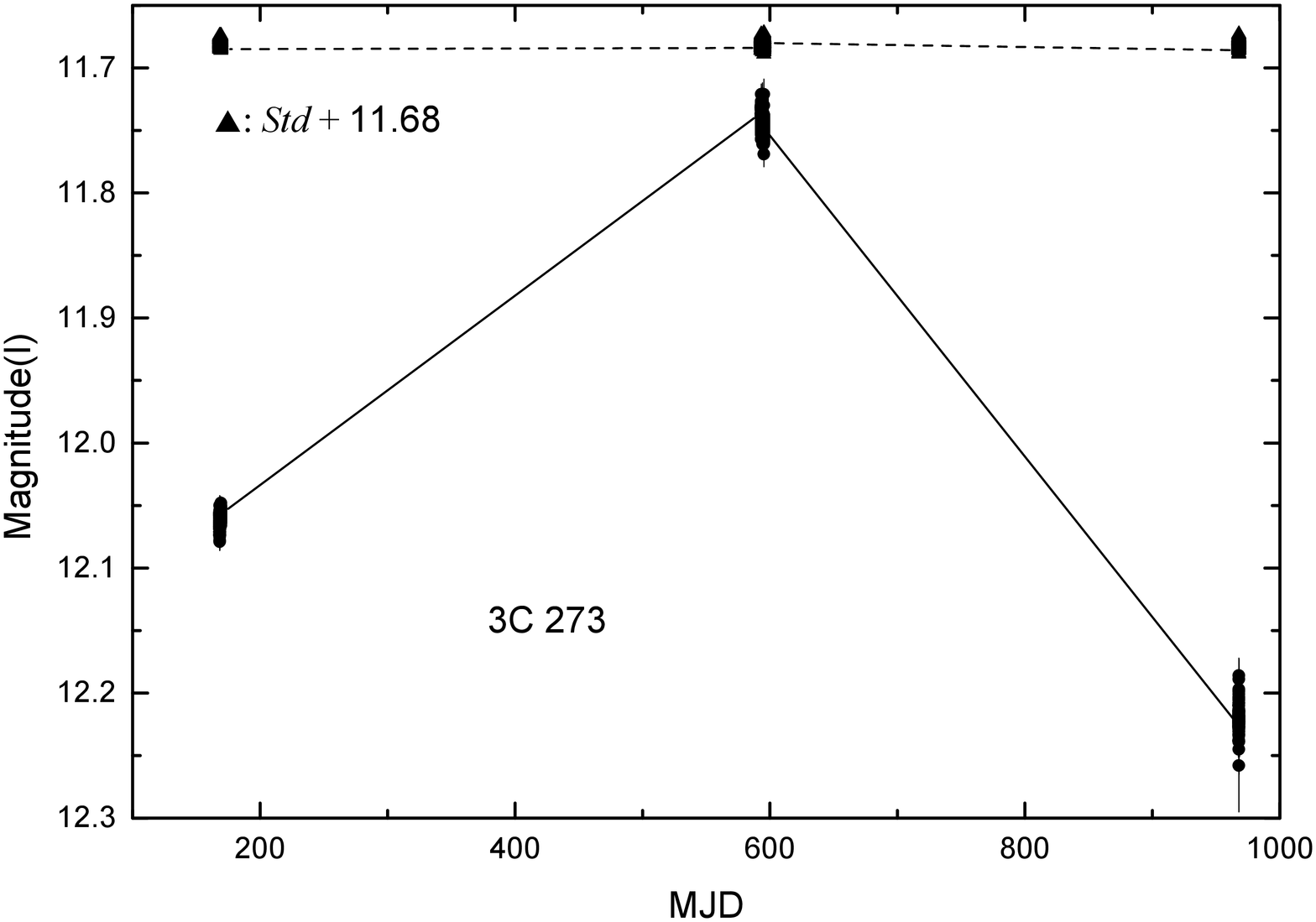}
  \includegraphics[angle=-90,scale=0.3]{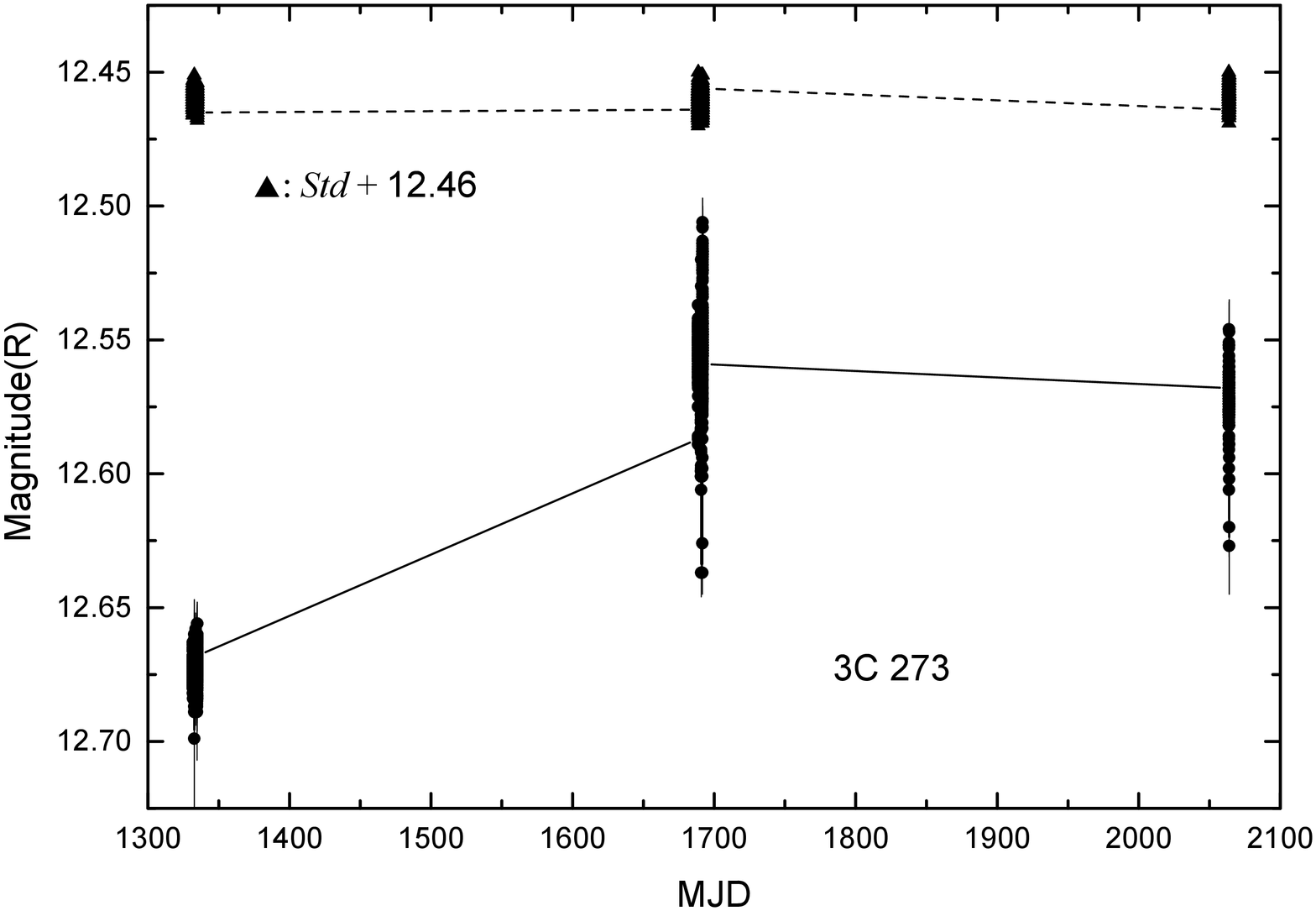}
 \end{center}
 \caption{Long term LCs of S5 0716+714 and 3C 273. Y-axes denote the apparent magnitudes. In each panel, the circles connected by the solid lines show the LC of blazar, and the triangles connected by the dashed lines denote the vertically moved $Std$ of the comparison stars, $Std^{\ast}$. The moved quantity is presented in each panel, such as $Std$+12.0.}
  \label{fig2}
\end{figure*}


\begin{figure*}
 \begin{center}
  \includegraphics[angle=-90,scale=0.25]{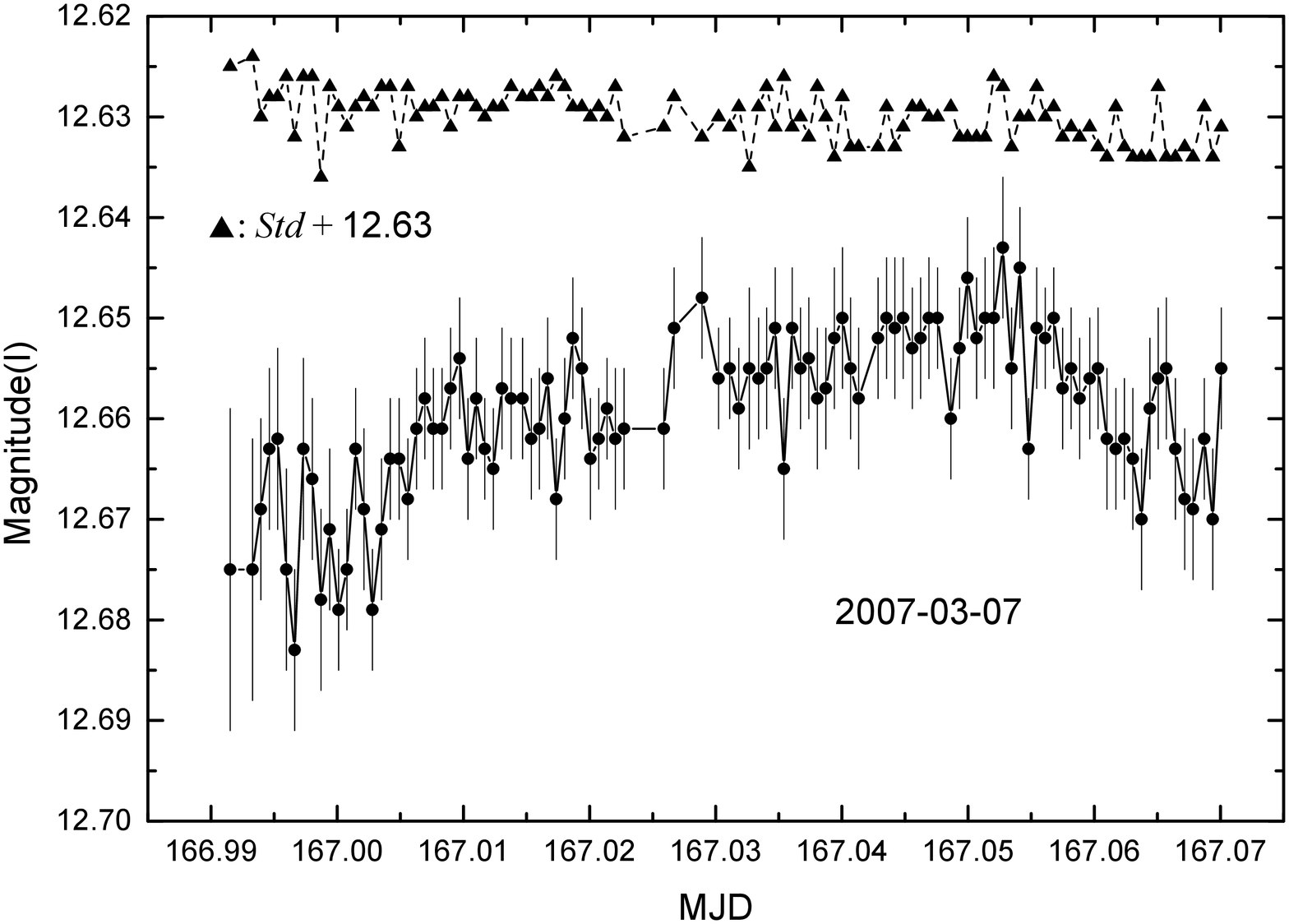}
  \includegraphics[angle=-90,scale=0.25]{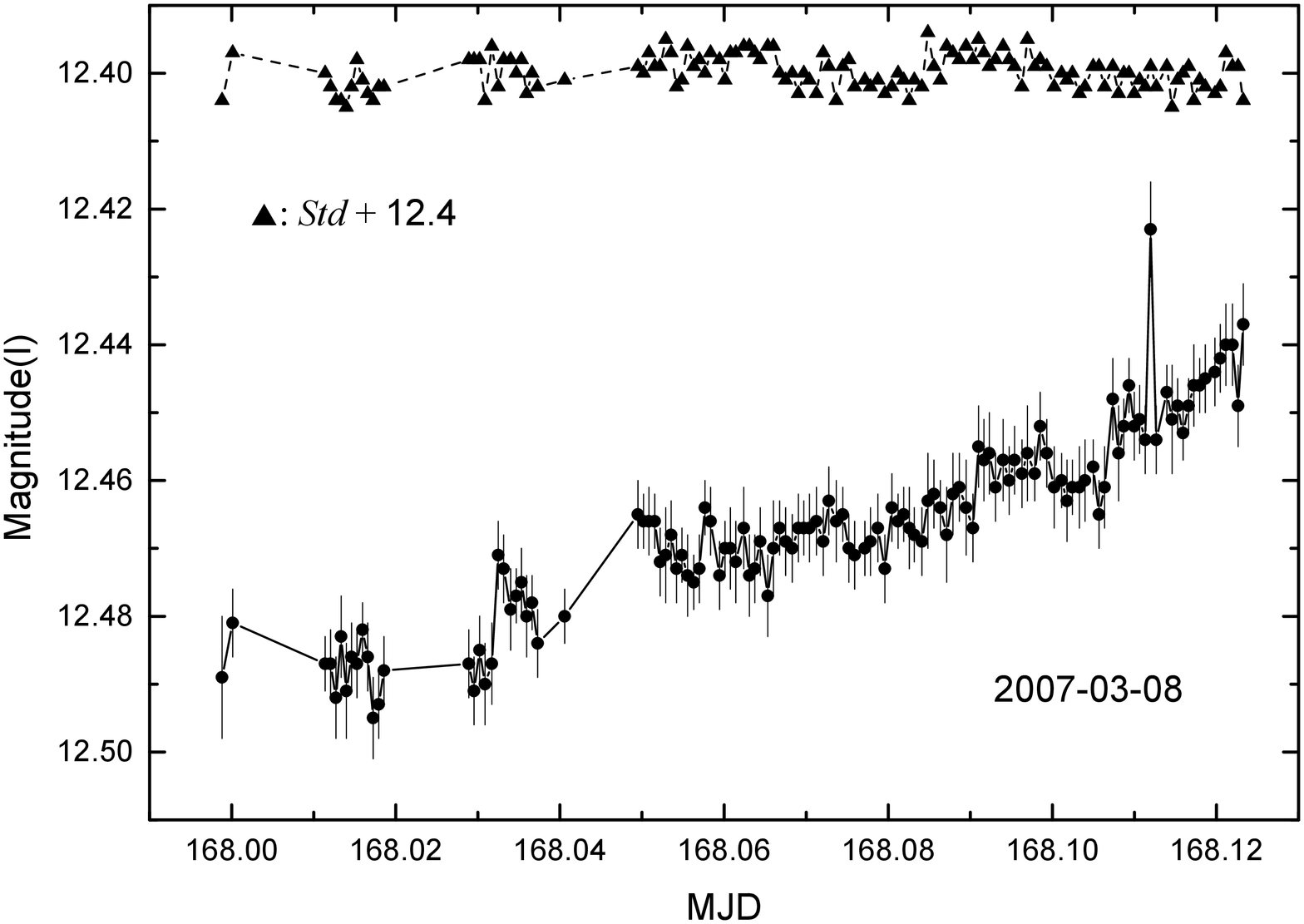}
  \includegraphics[angle=-90,scale=0.25]{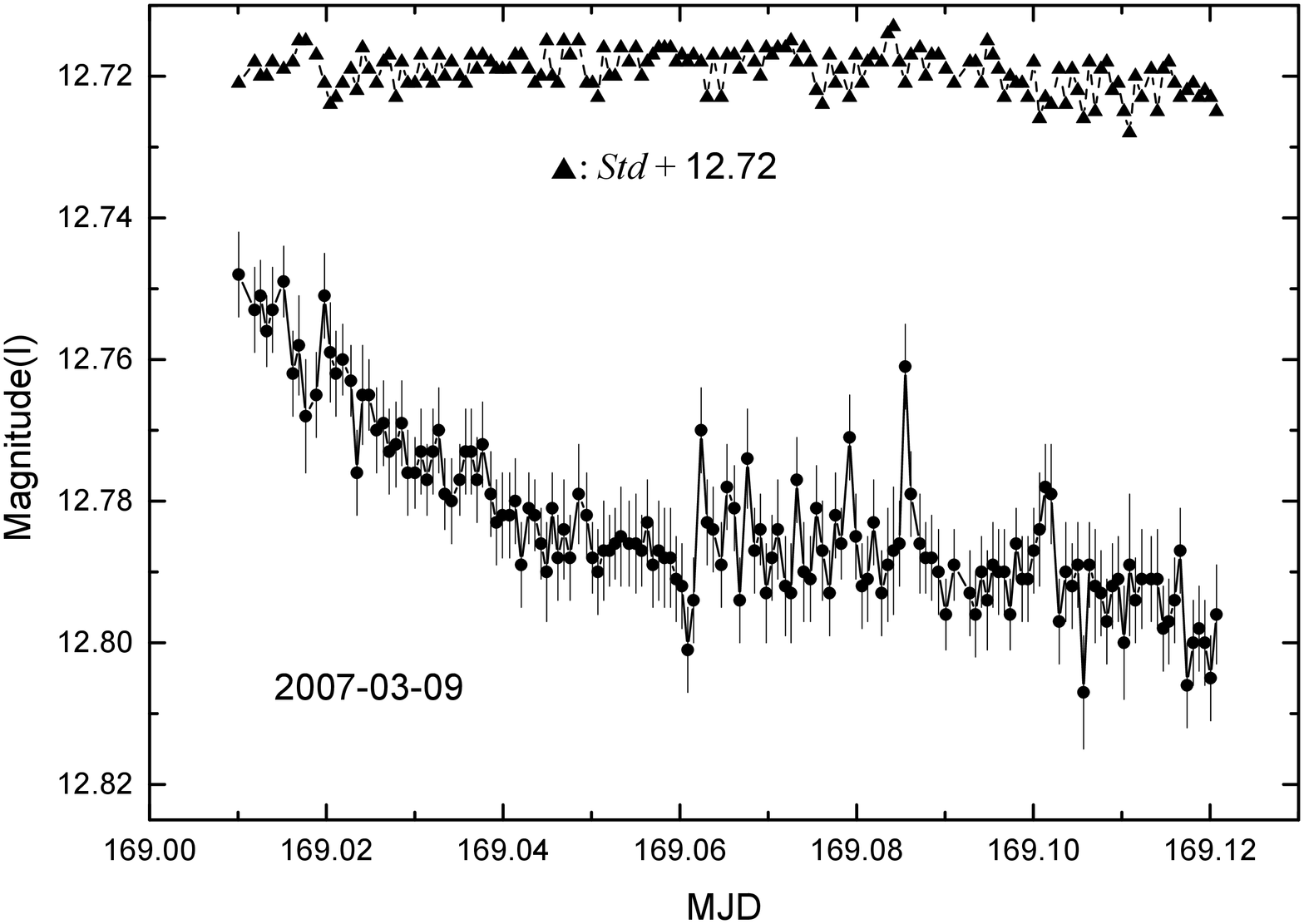}
  \includegraphics[angle=-90,scale=0.25]{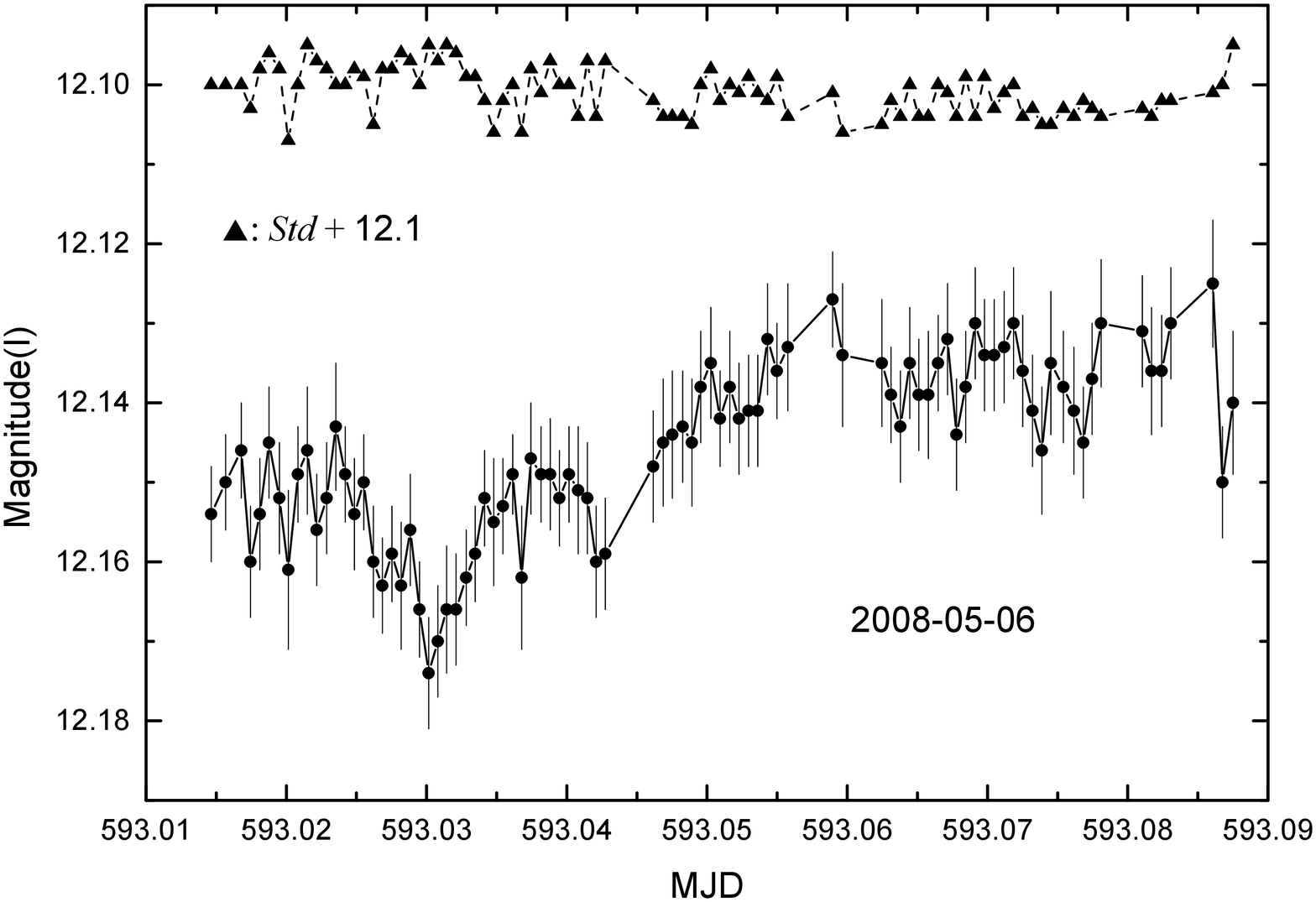}
  \includegraphics[angle=-90,scale=0.25]{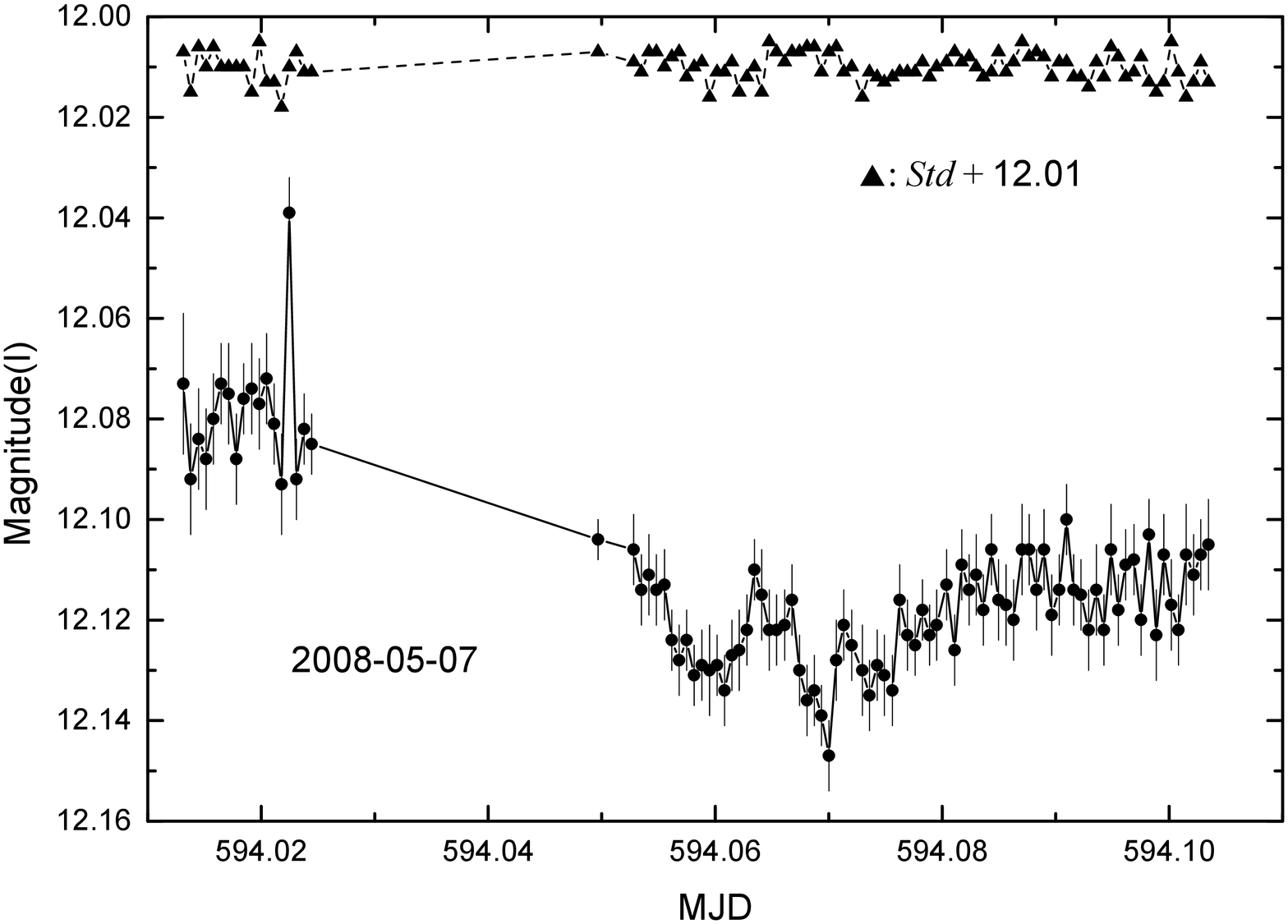}
  \includegraphics[angle=-90,scale=0.25]{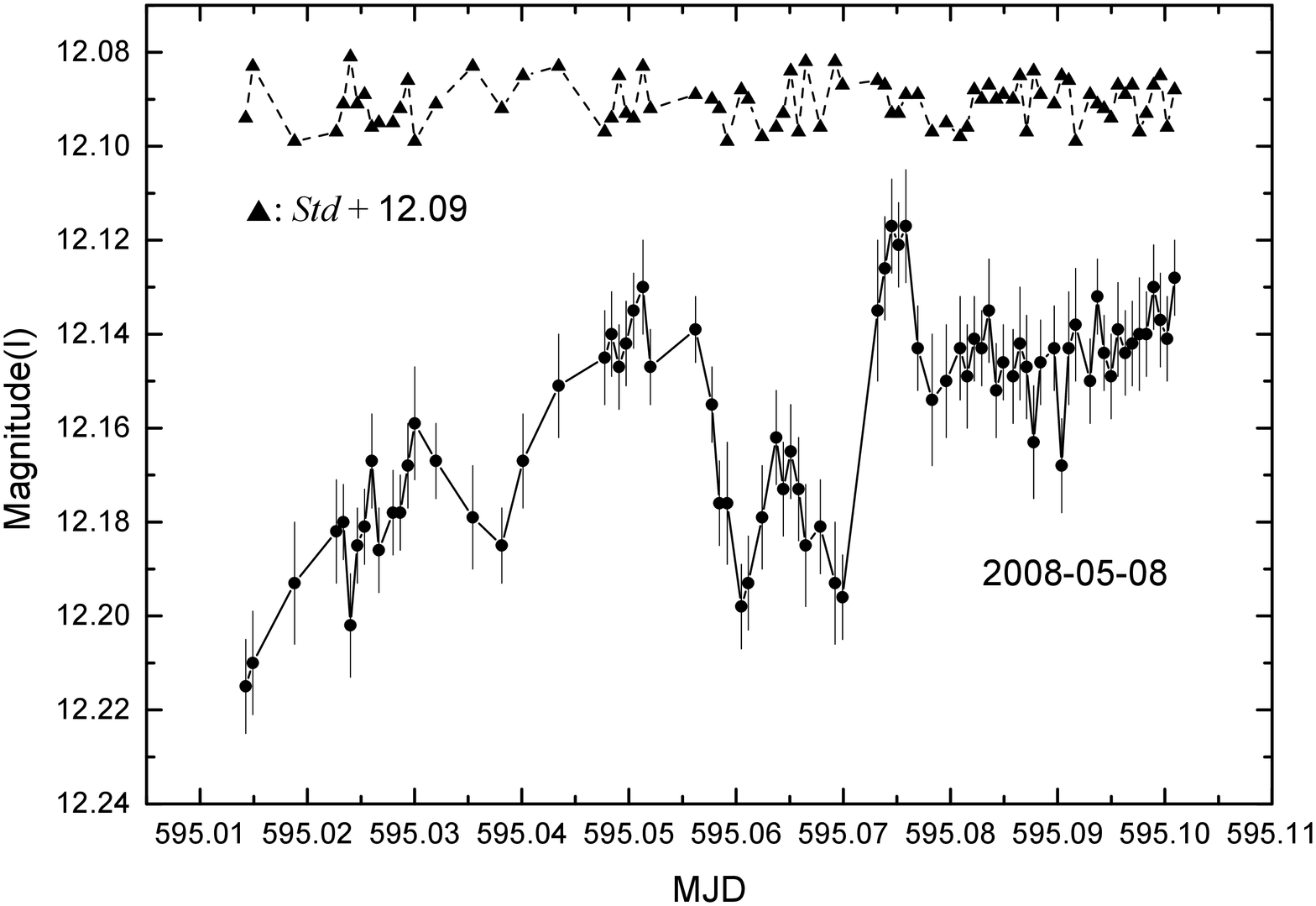}
 \end{center}
 \caption{LCs of S5 0716+714 (same symbols as Figure 2). Y-axes denote the apparent magnitudes.}
  \label{fig3}
\end{figure*}


\begin{figure*}
 \begin{center}
  \includegraphics[angle=-90,scale=0.25]{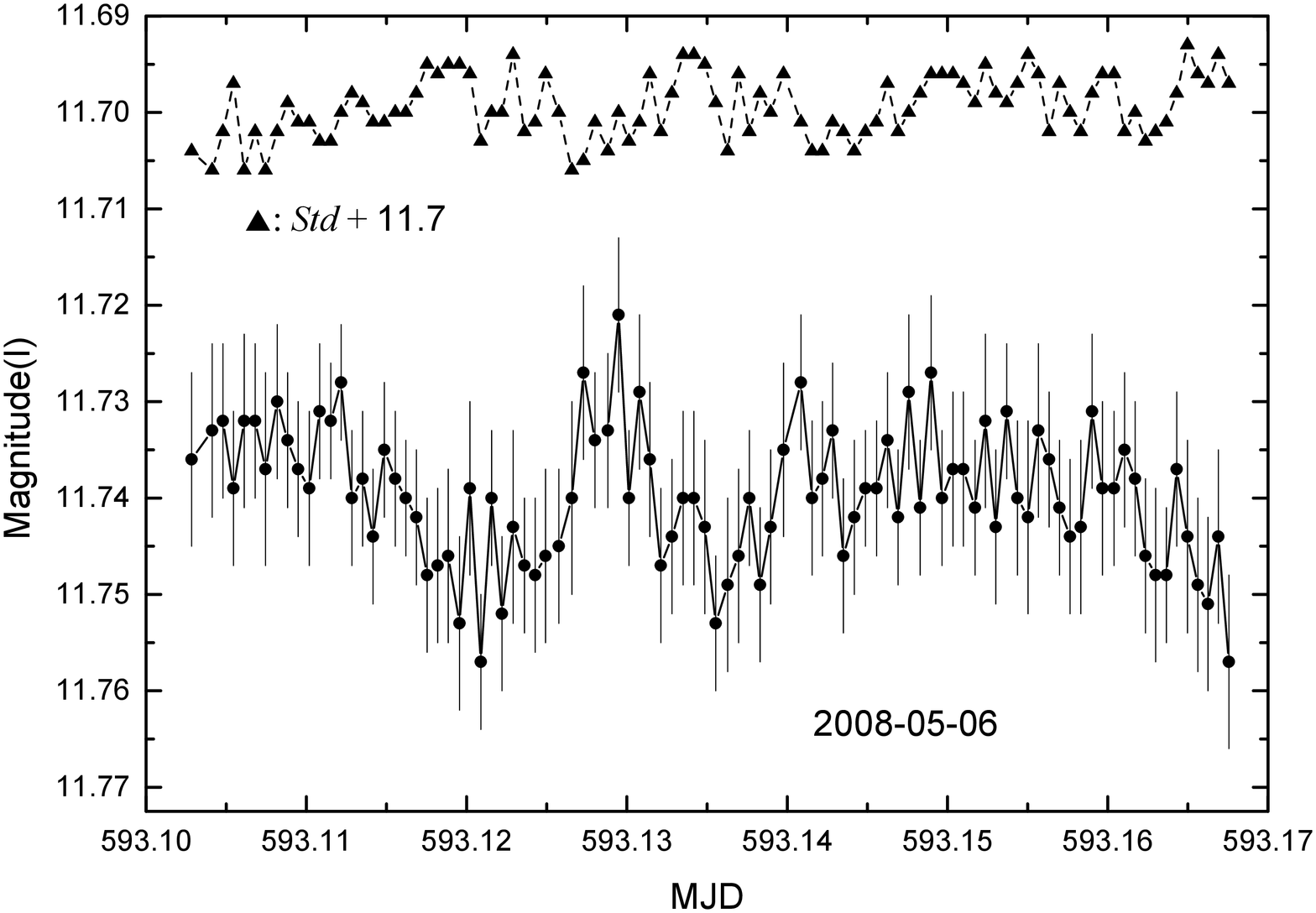}
  \includegraphics[angle=-90,scale=0.25]{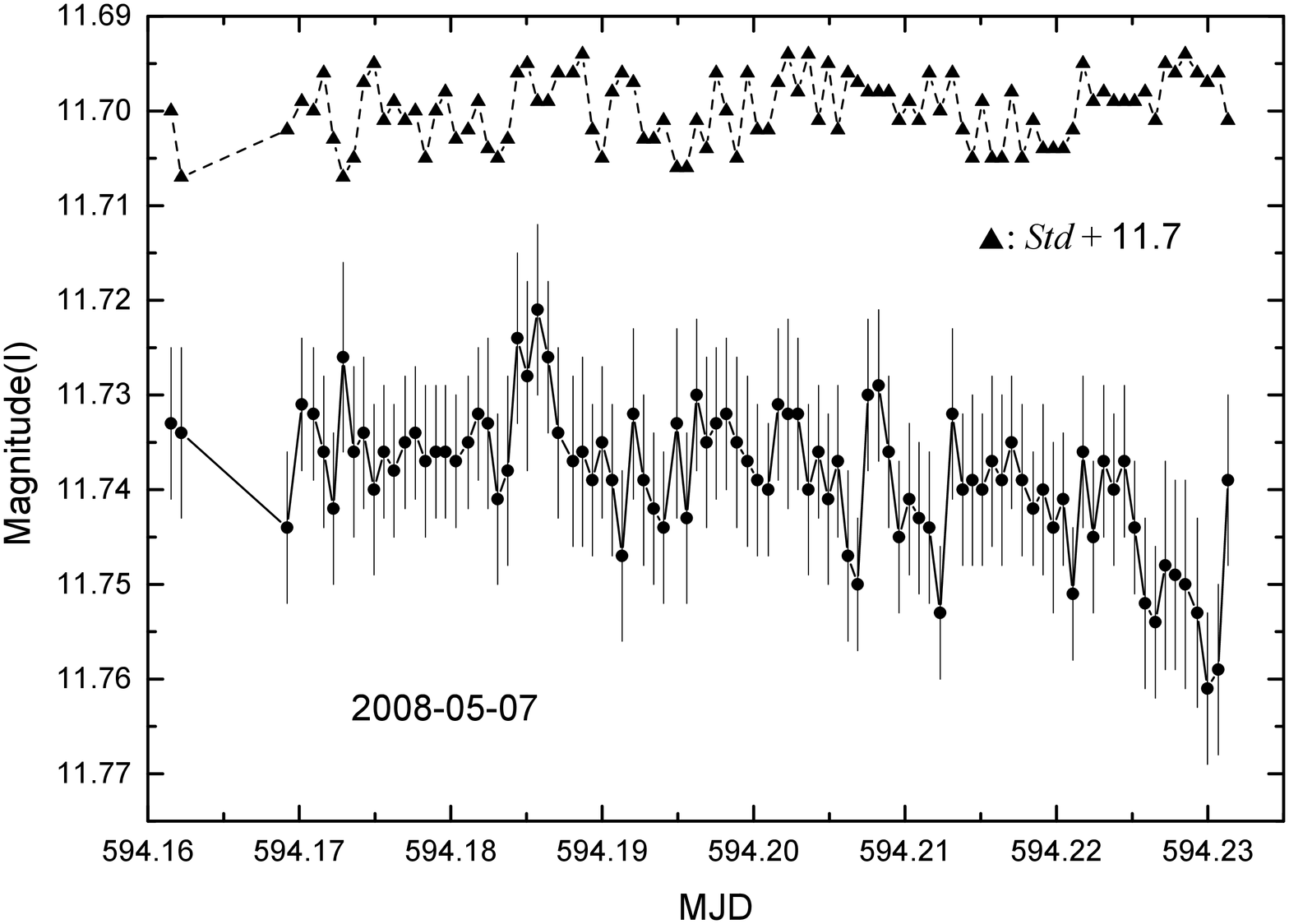}
  \includegraphics[angle=-90,scale=0.25]{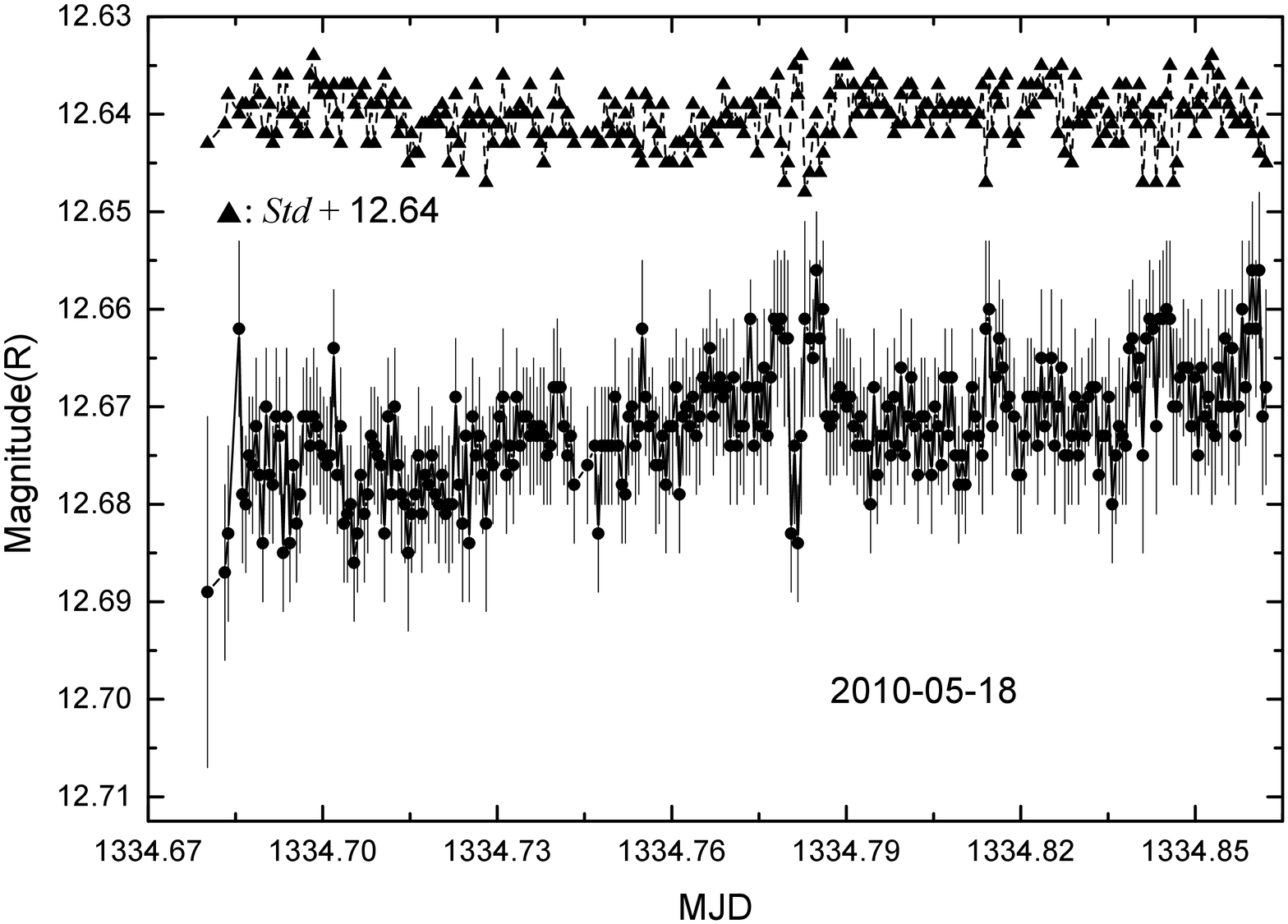}
  \includegraphics[angle=-90,scale=0.25]{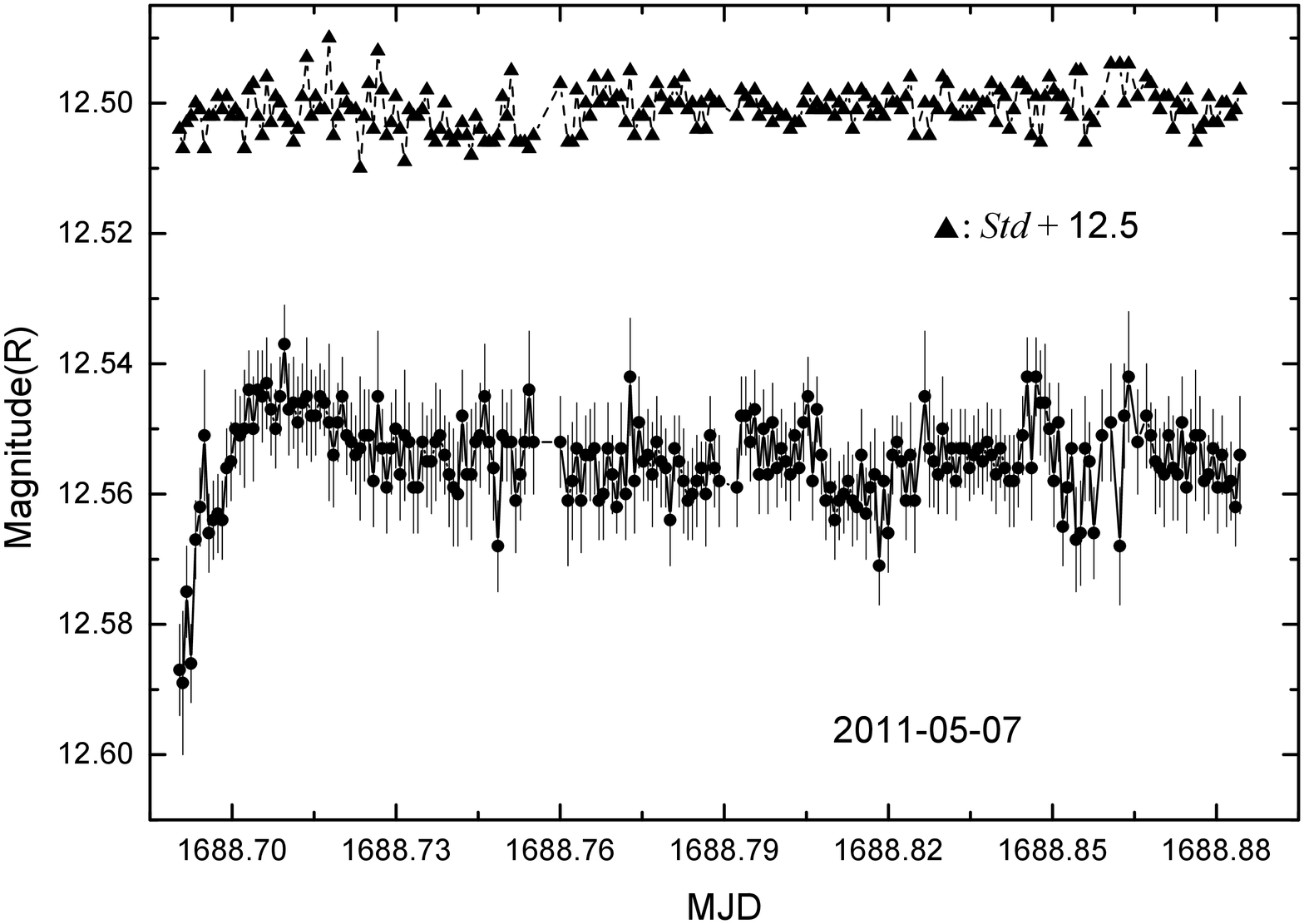}
  \includegraphics[angle=-90,scale=0.25]{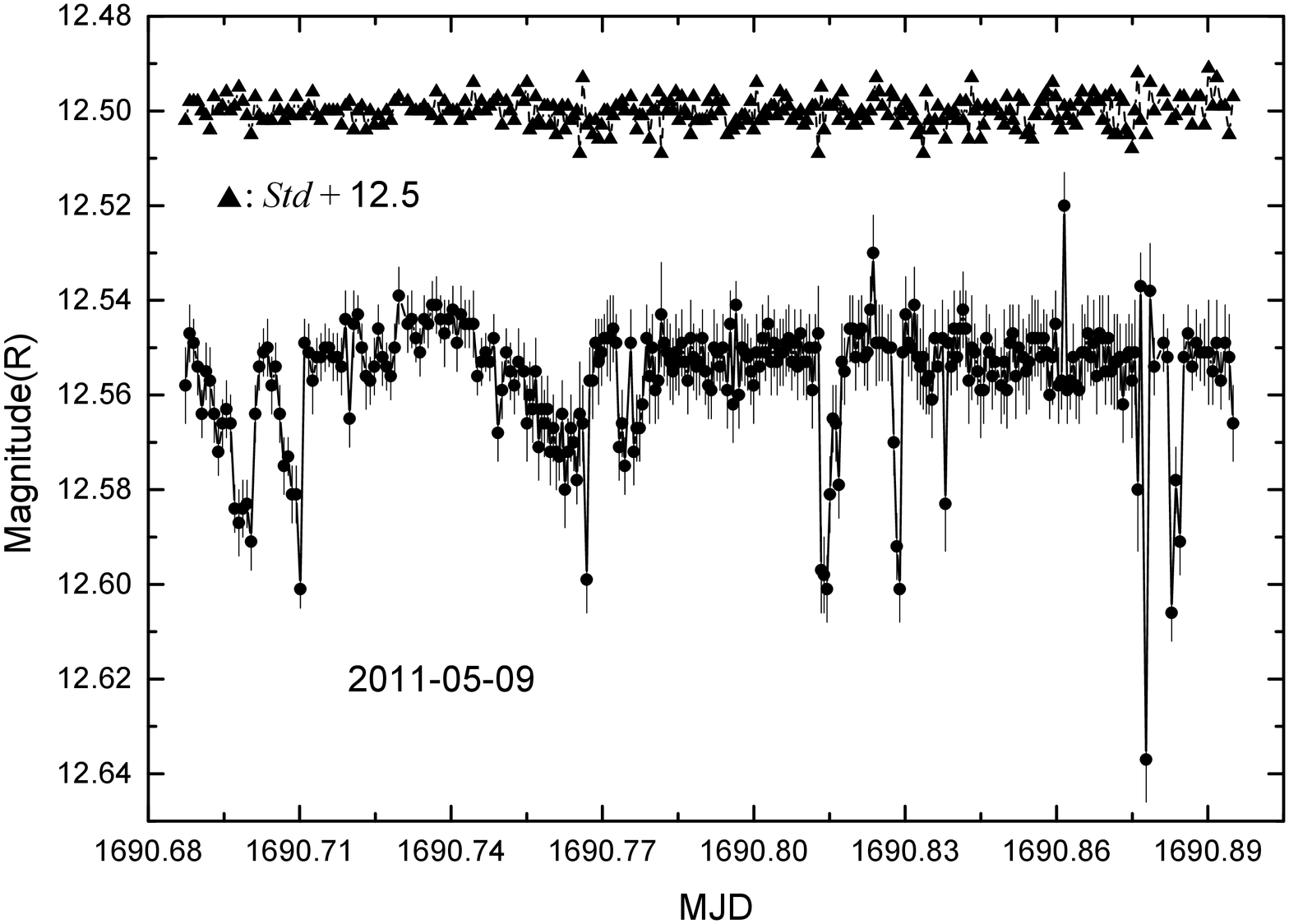}
  \includegraphics[angle=-90,scale=0.25]{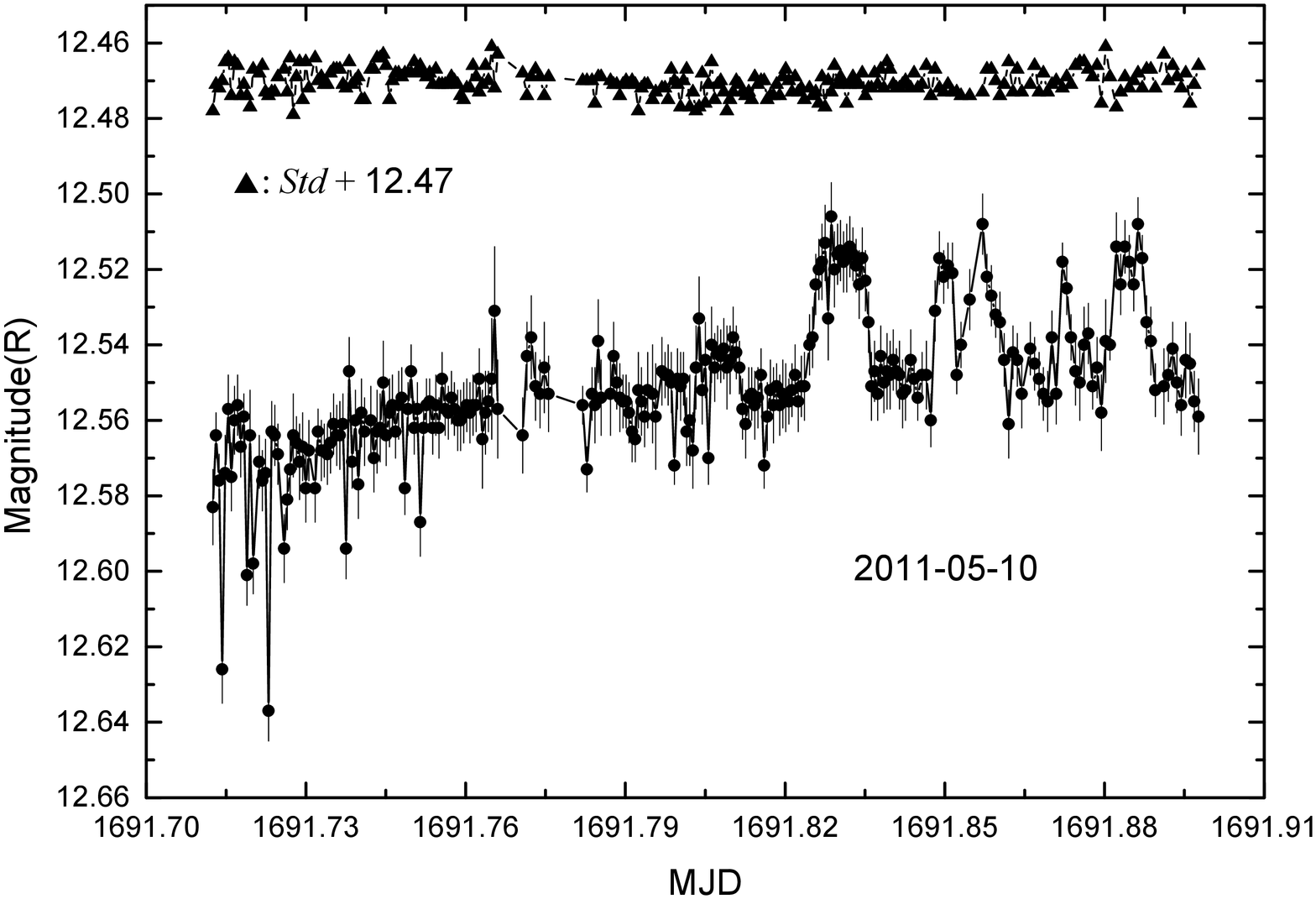}
 \end{center}
 \caption{LCs of 3C 273 (same symbols as Figure 2). Y-axes denote the apparent magnitudes.}
  \label{fig4}
\end{figure*}


\begin{figure*}
 \begin{center}
  \includegraphics[angle=0,scale=0.25]{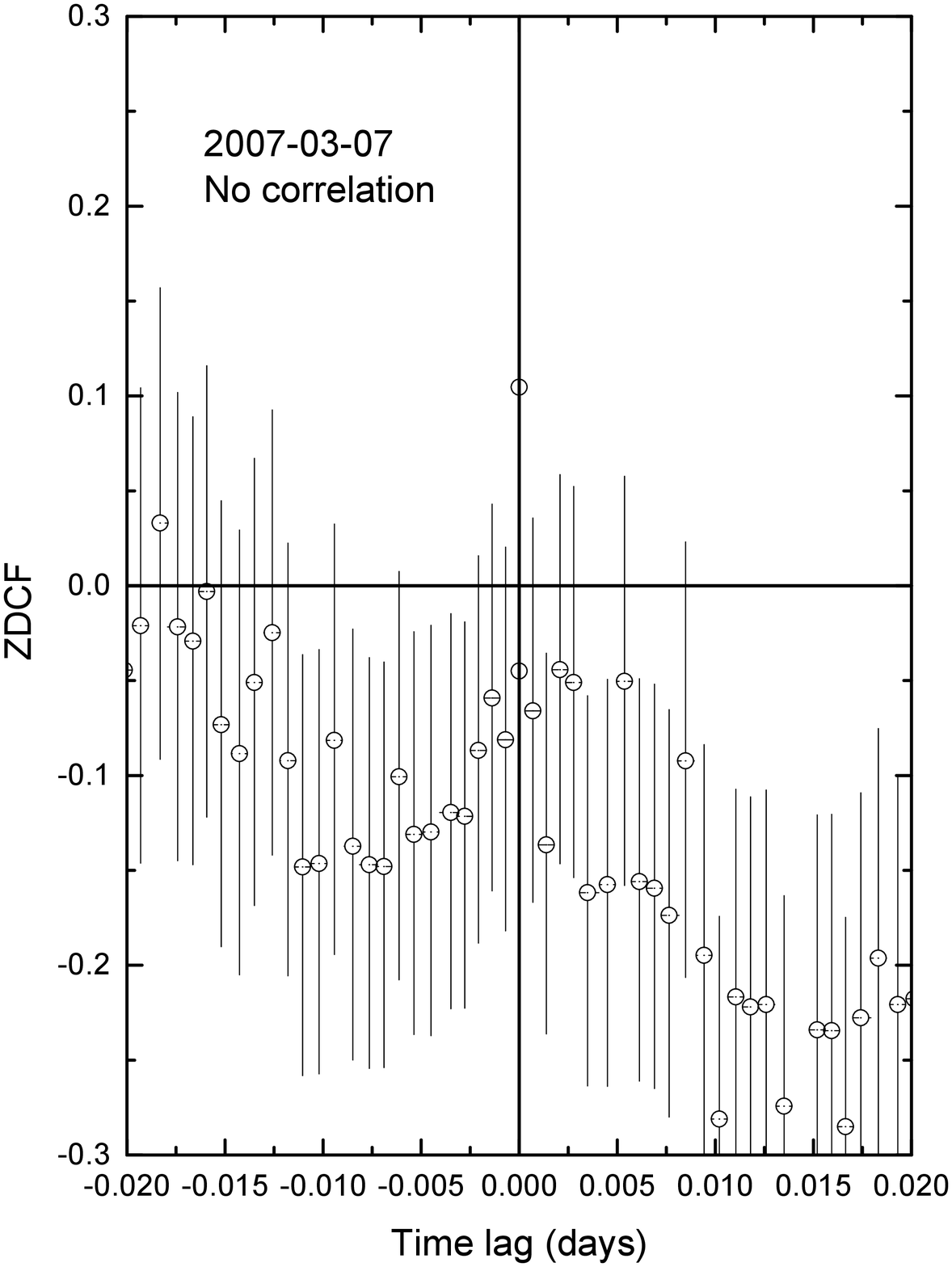}
  \includegraphics[angle=0,scale=0.25]{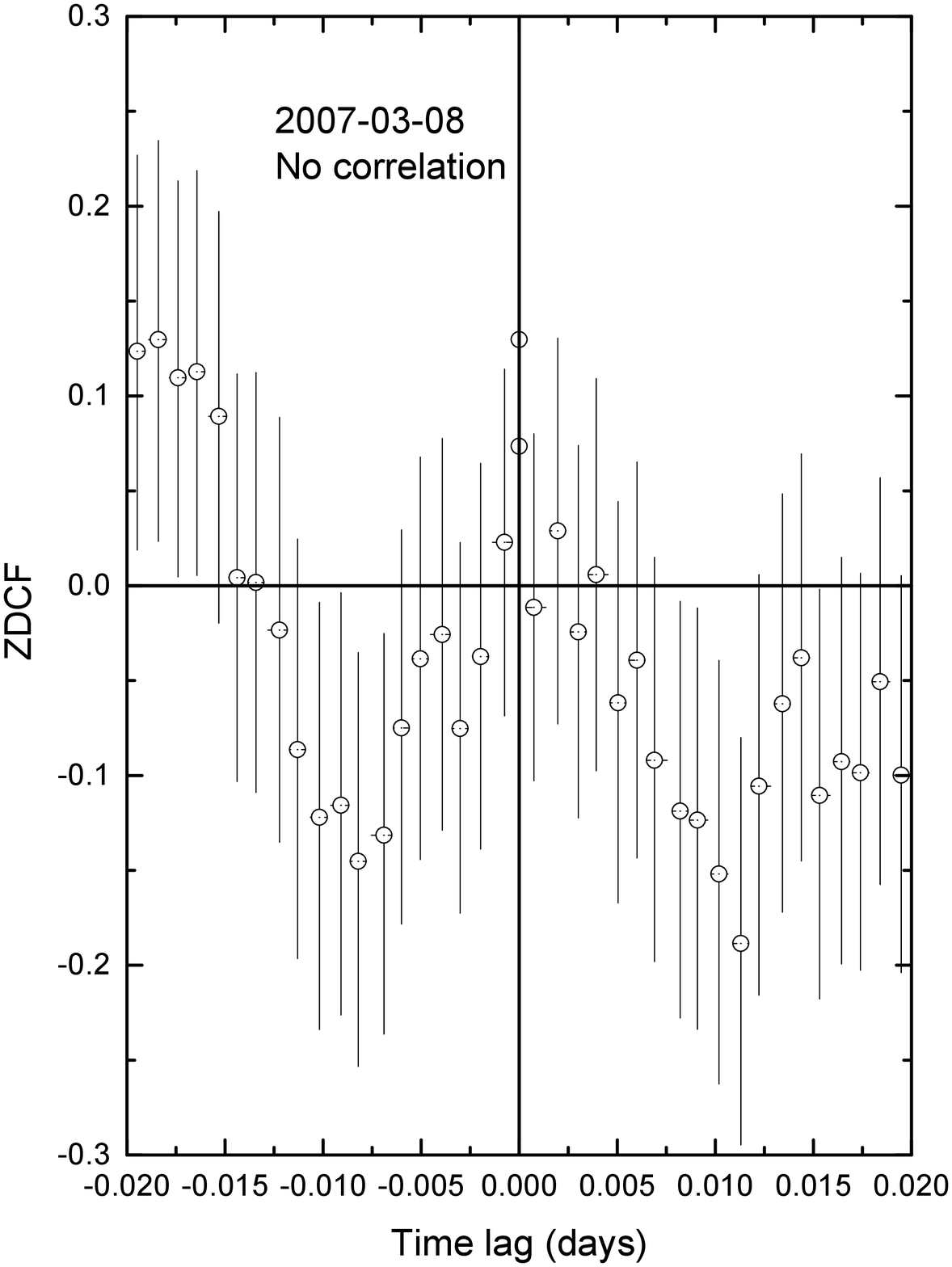}
  \includegraphics[angle=0,scale=0.25]{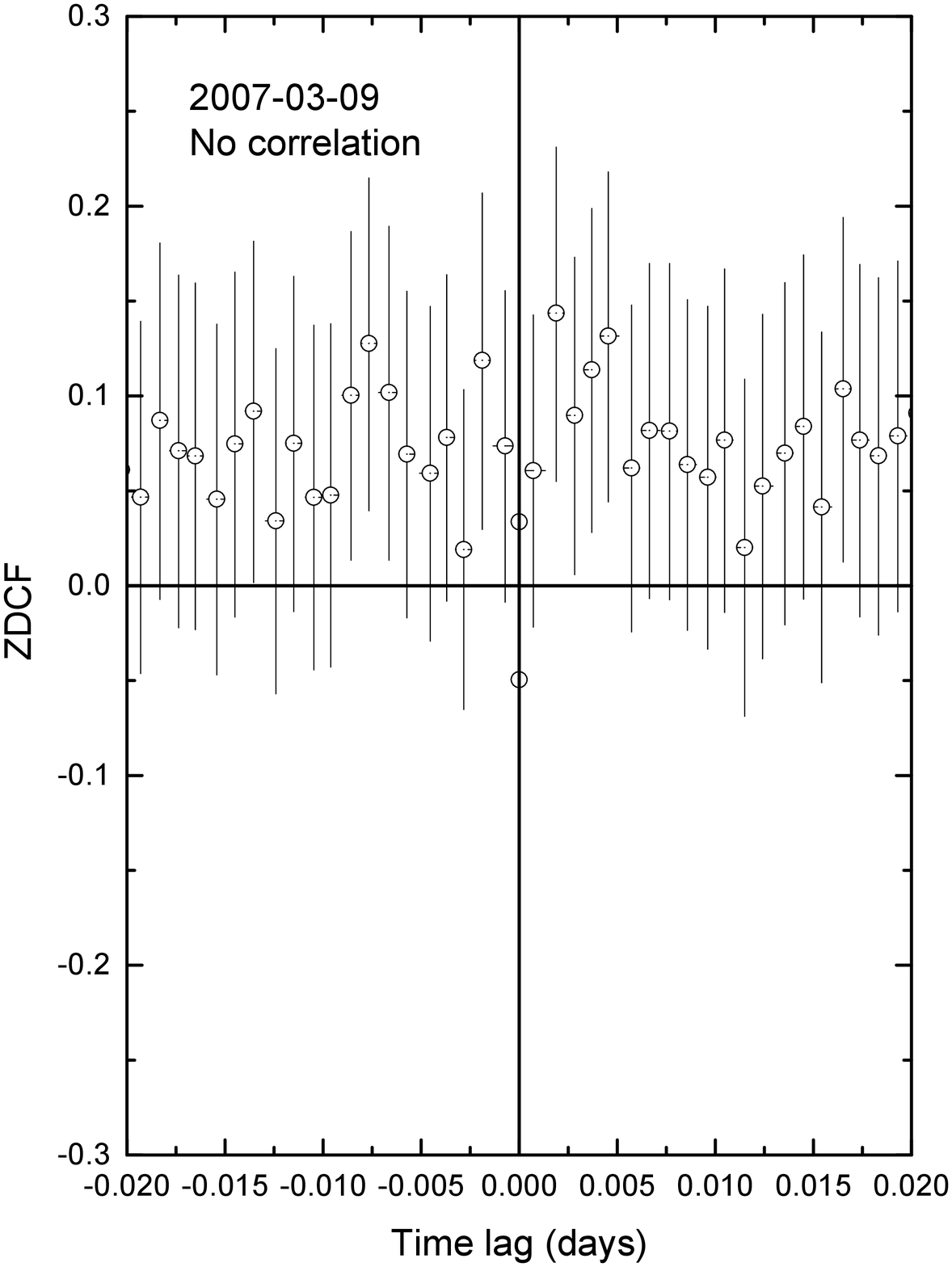}
  \includegraphics[angle=0,scale=0.25]{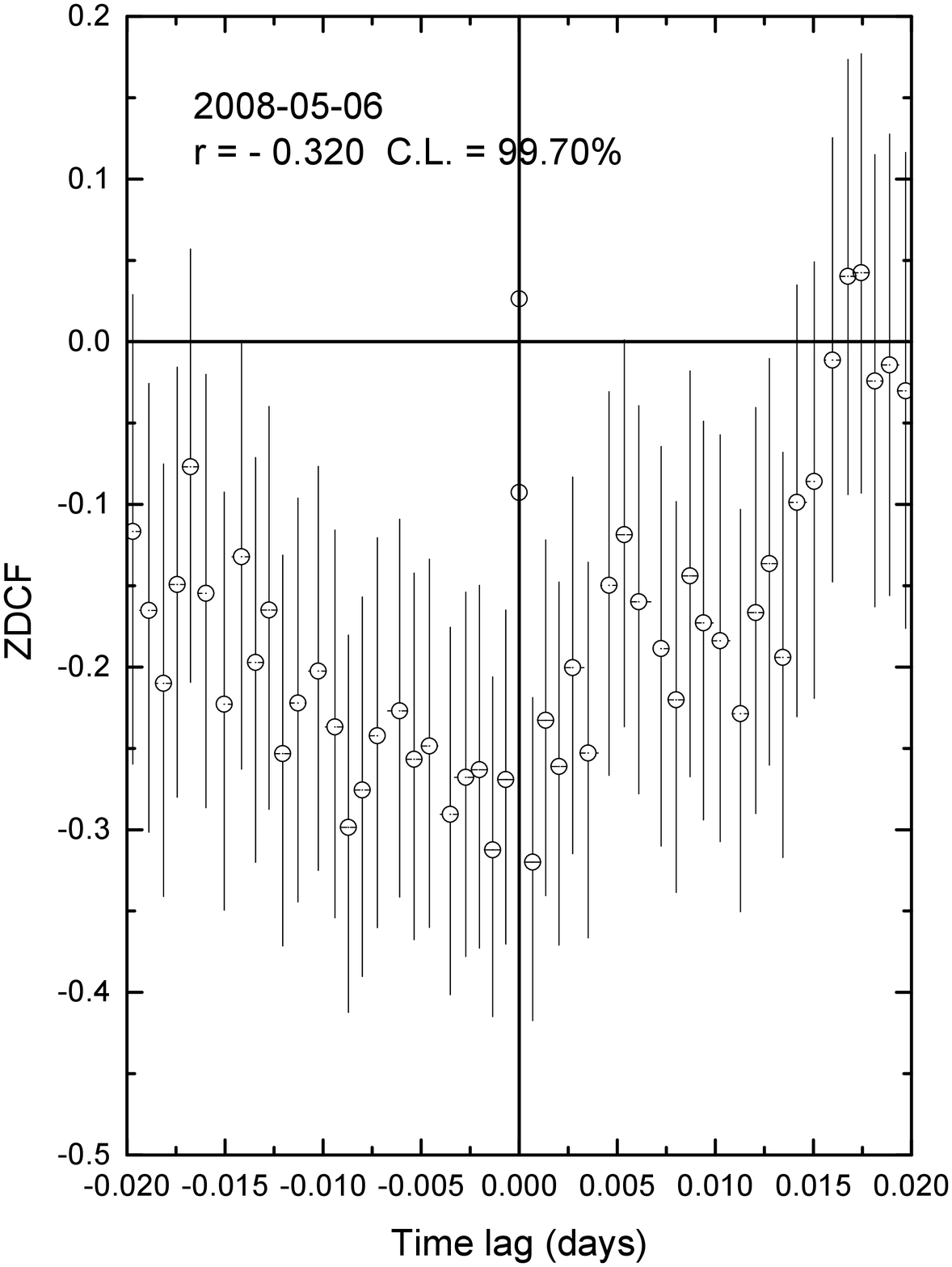}
  \includegraphics[angle=0,scale=0.25]{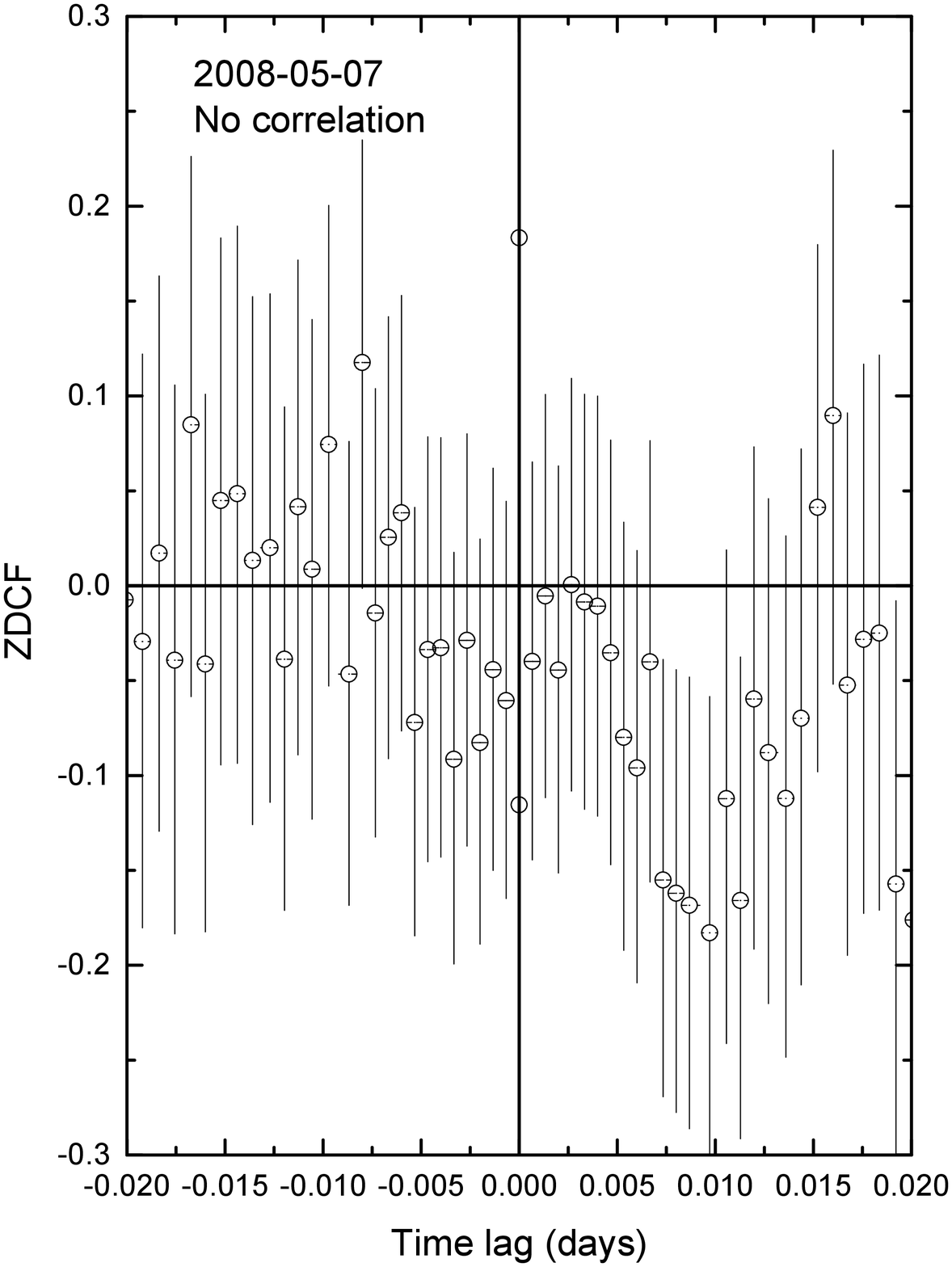}
  \includegraphics[angle=0,scale=0.25]{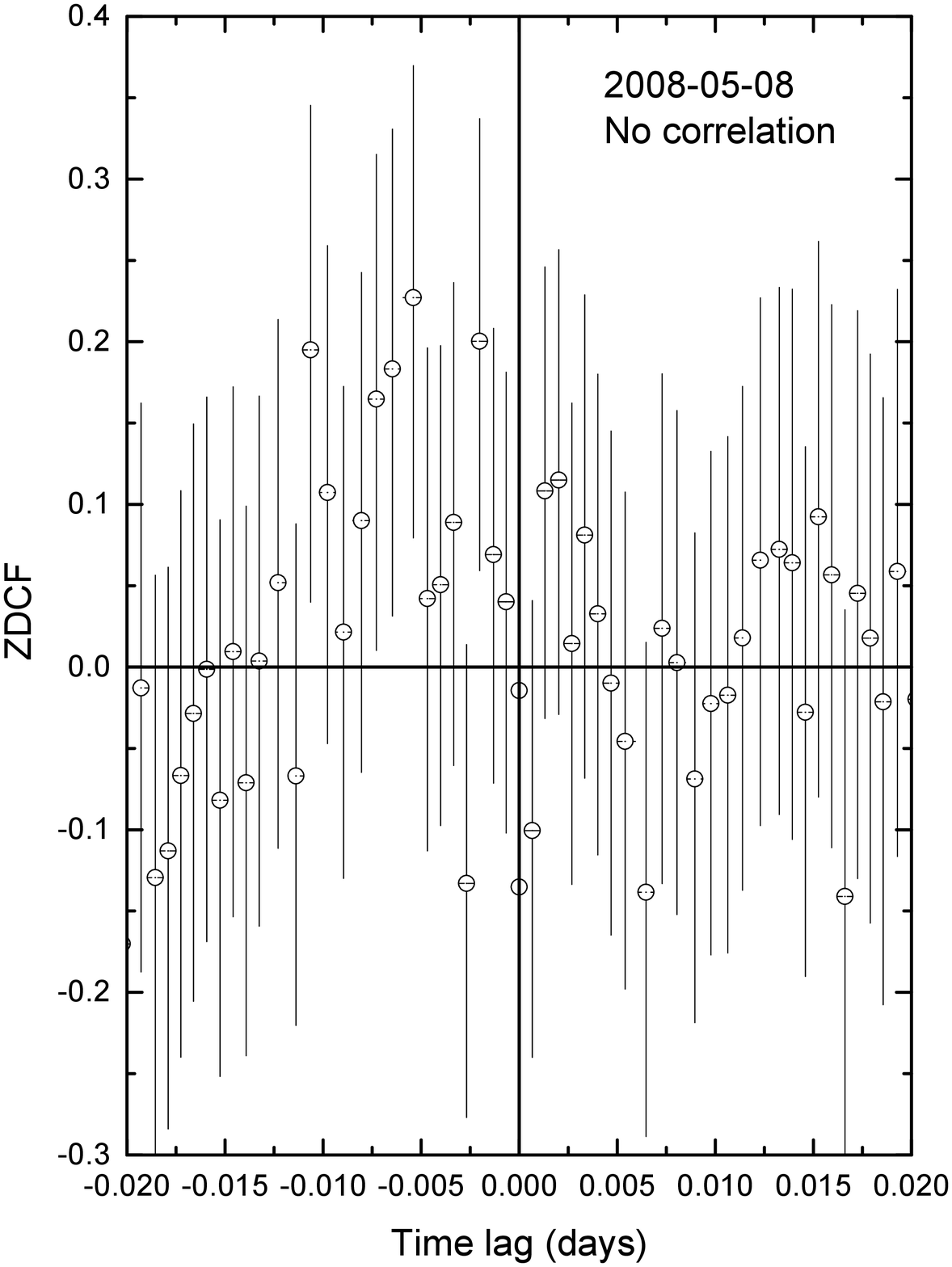}
 \end{center}
 \caption{ZDCFs calculated from the LCs in Figure 3 for S5 0716+714 and $Std^{\ast}$. }
  \label{fig5}
\end{figure*}


\begin{figure*}
 \begin{center}
  \includegraphics[angle=0,scale=0.25]{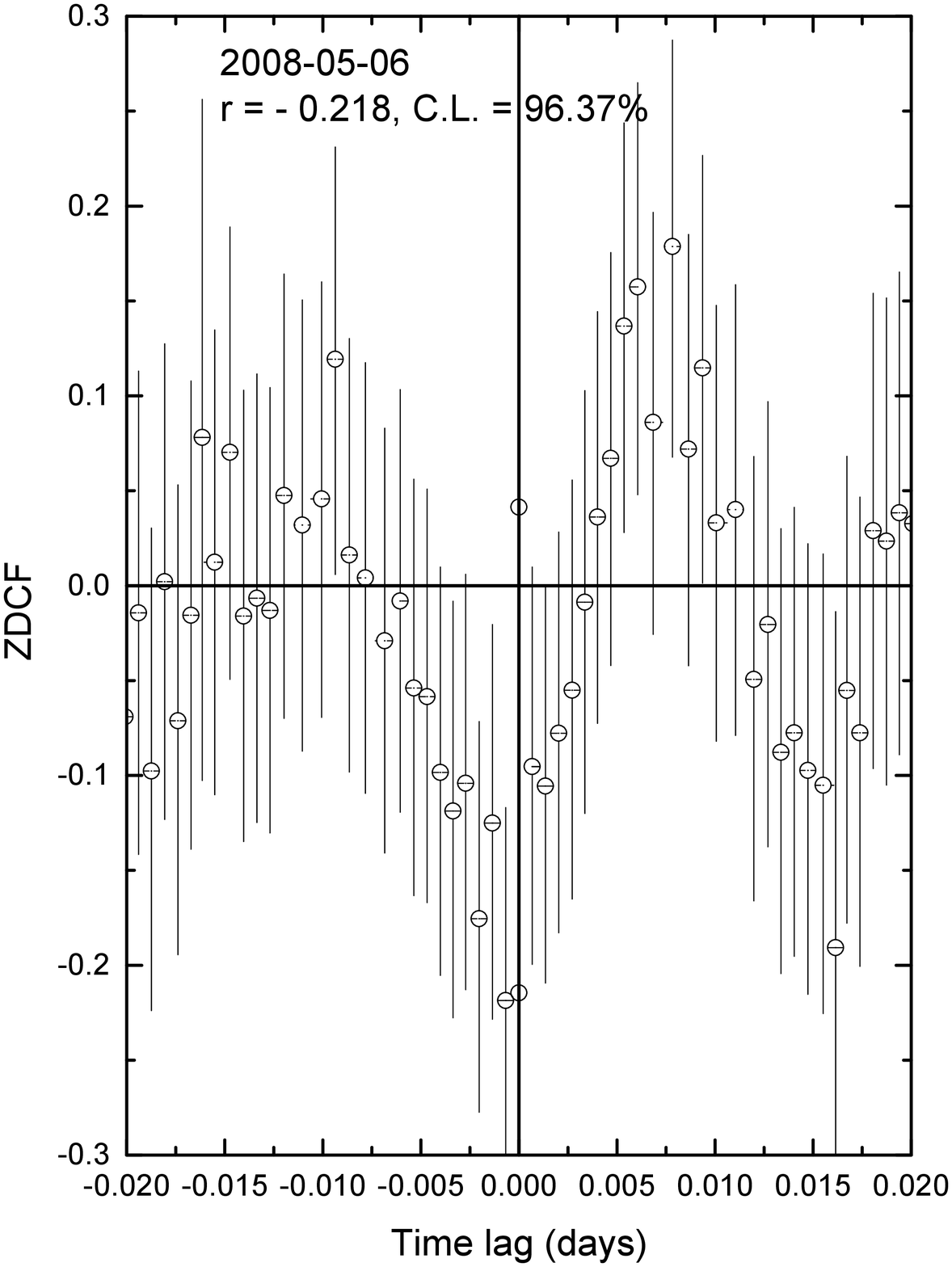}
  \includegraphics[angle=0,scale=0.25]{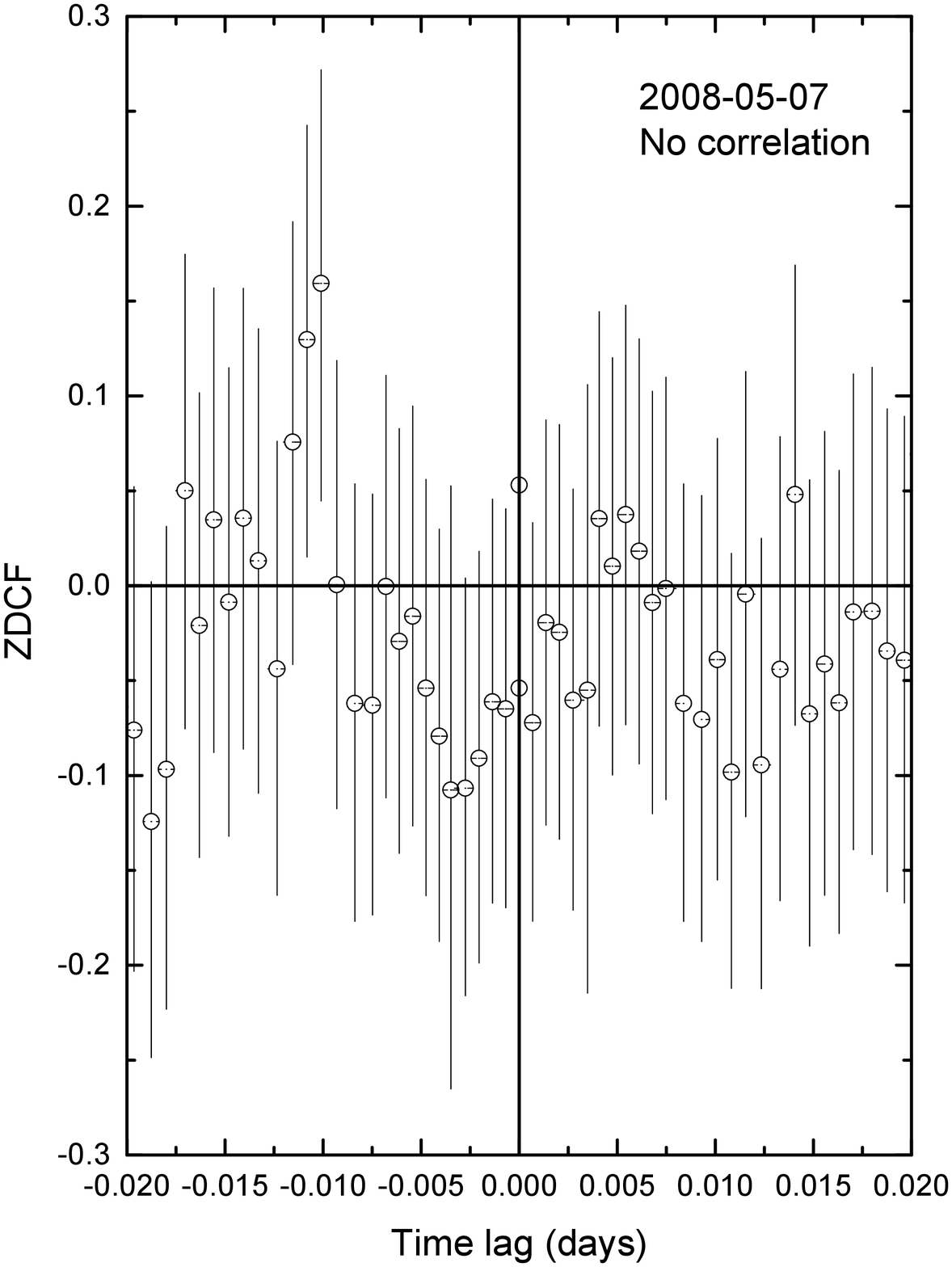}
  \includegraphics[angle=0,scale=0.25]{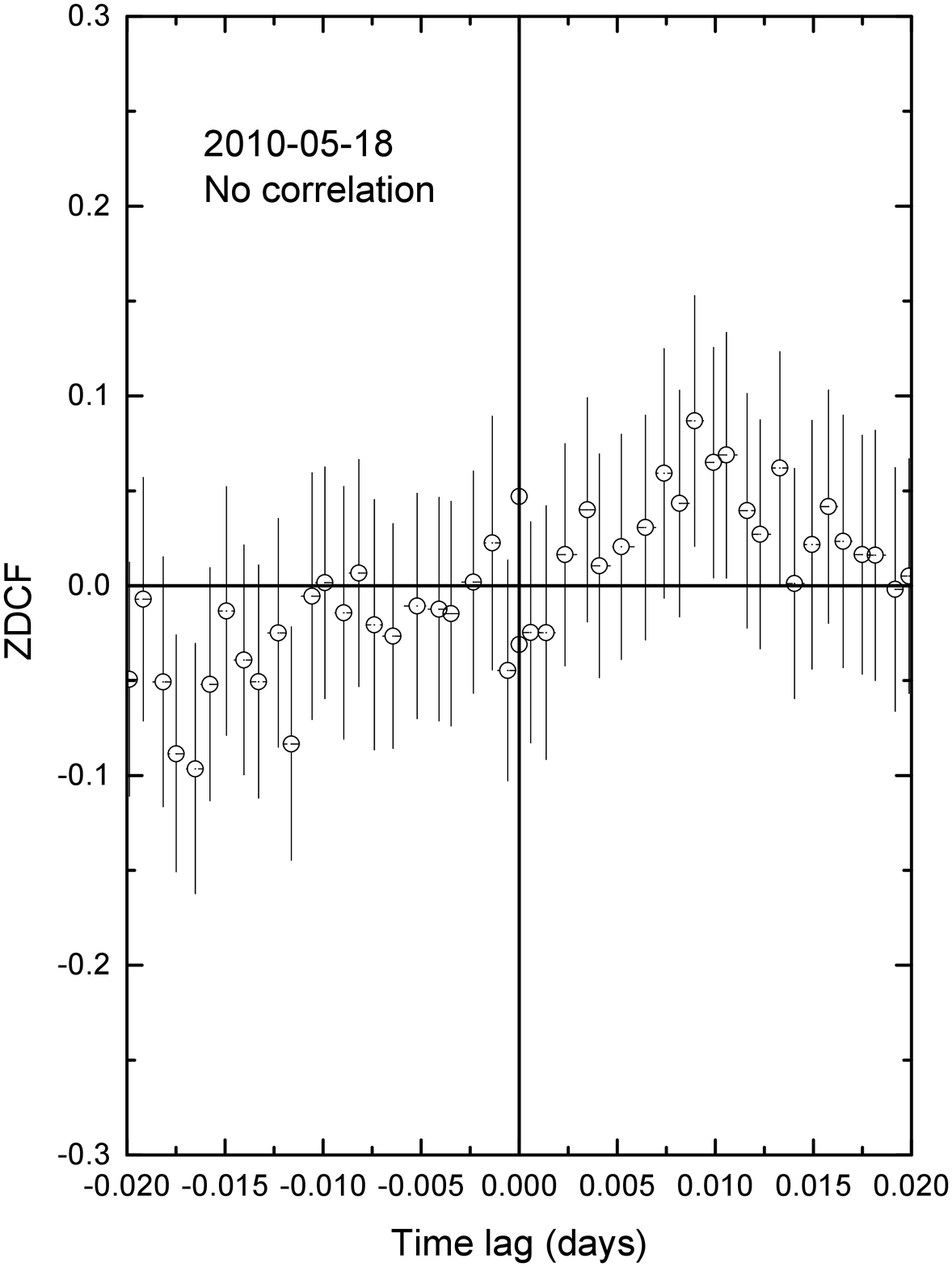}
  \includegraphics[angle=0,scale=0.25]{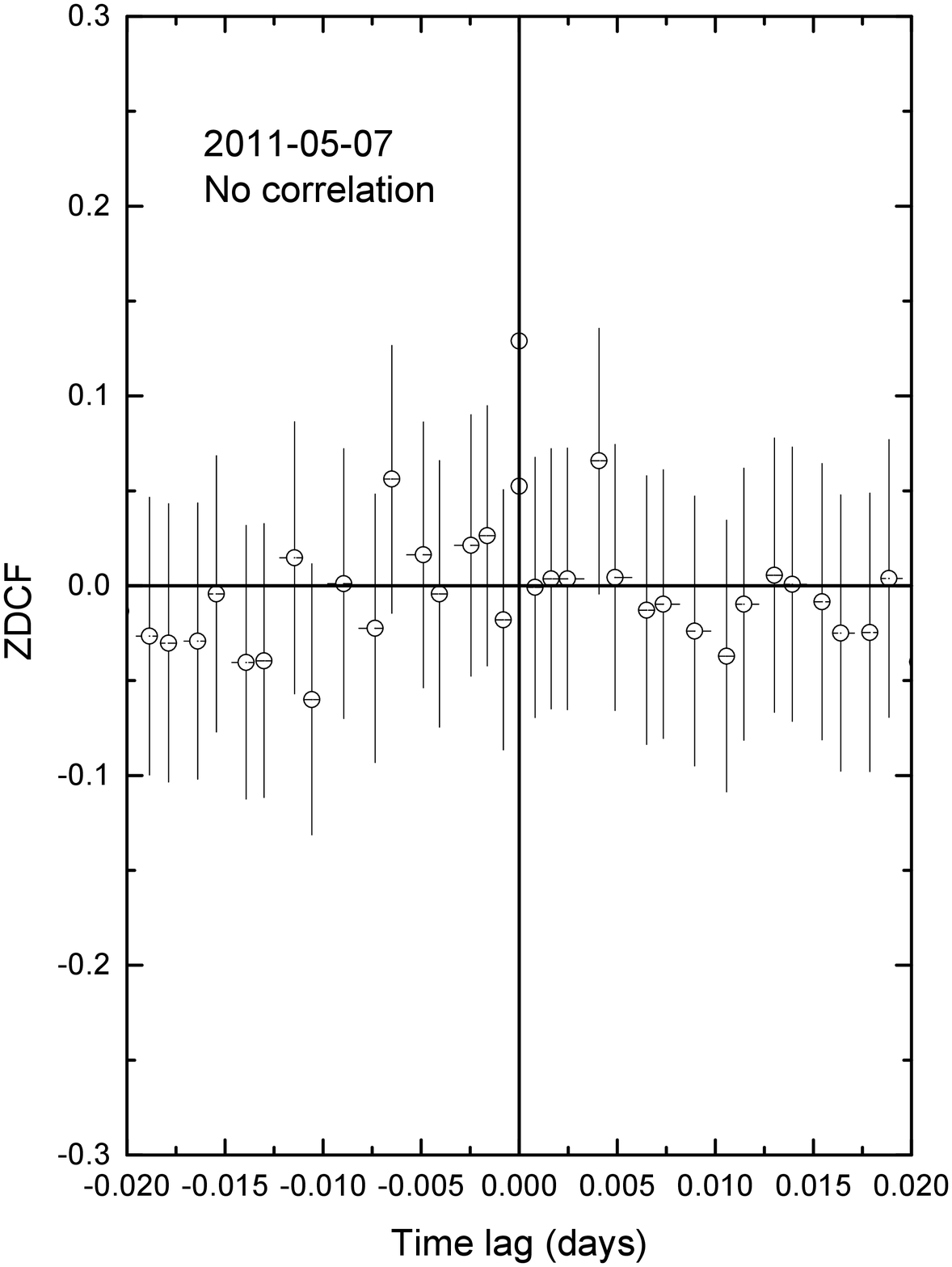}
  \includegraphics[angle=0,scale=0.25]{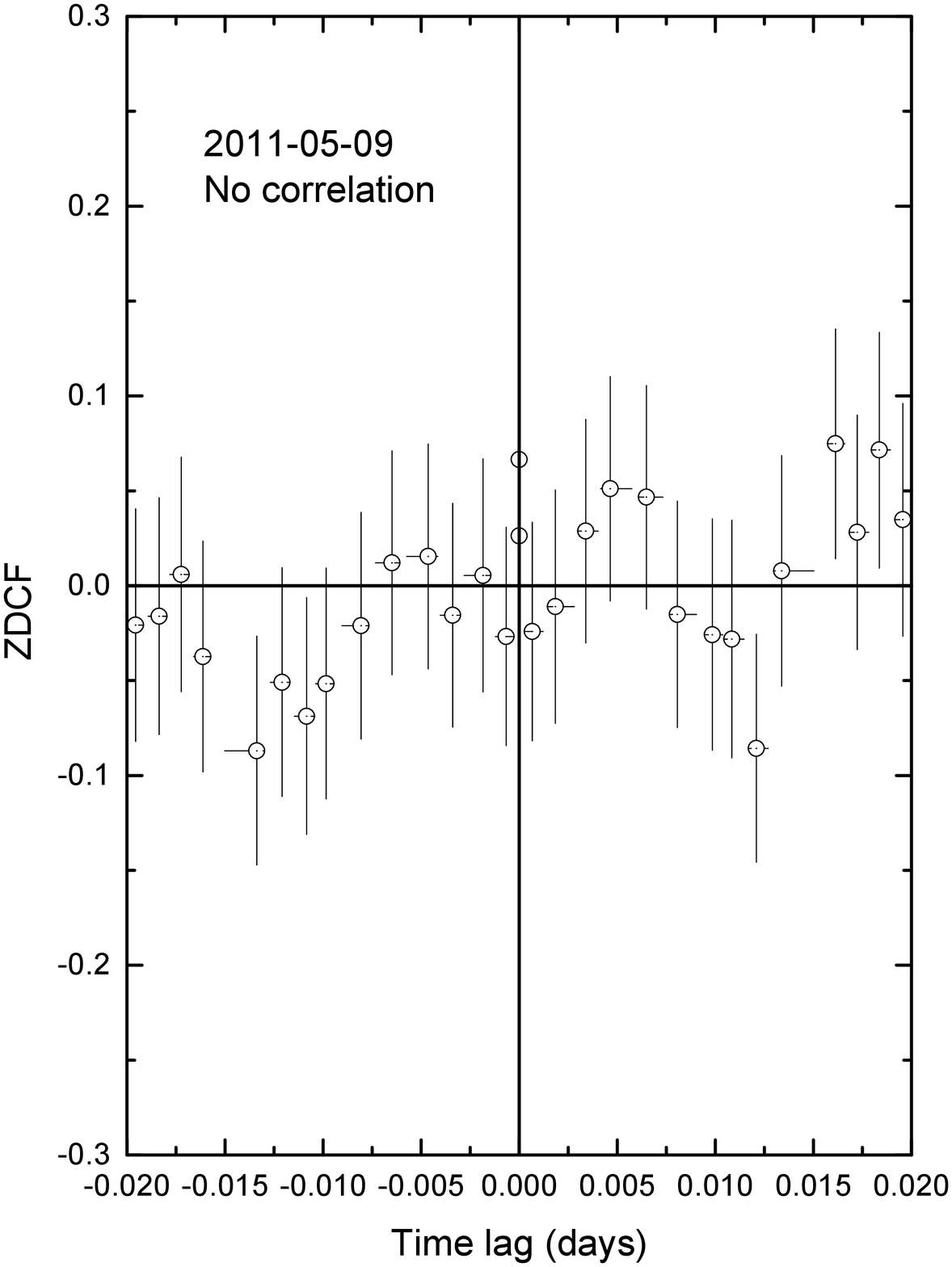}
  \includegraphics[angle=0,scale=0.25]{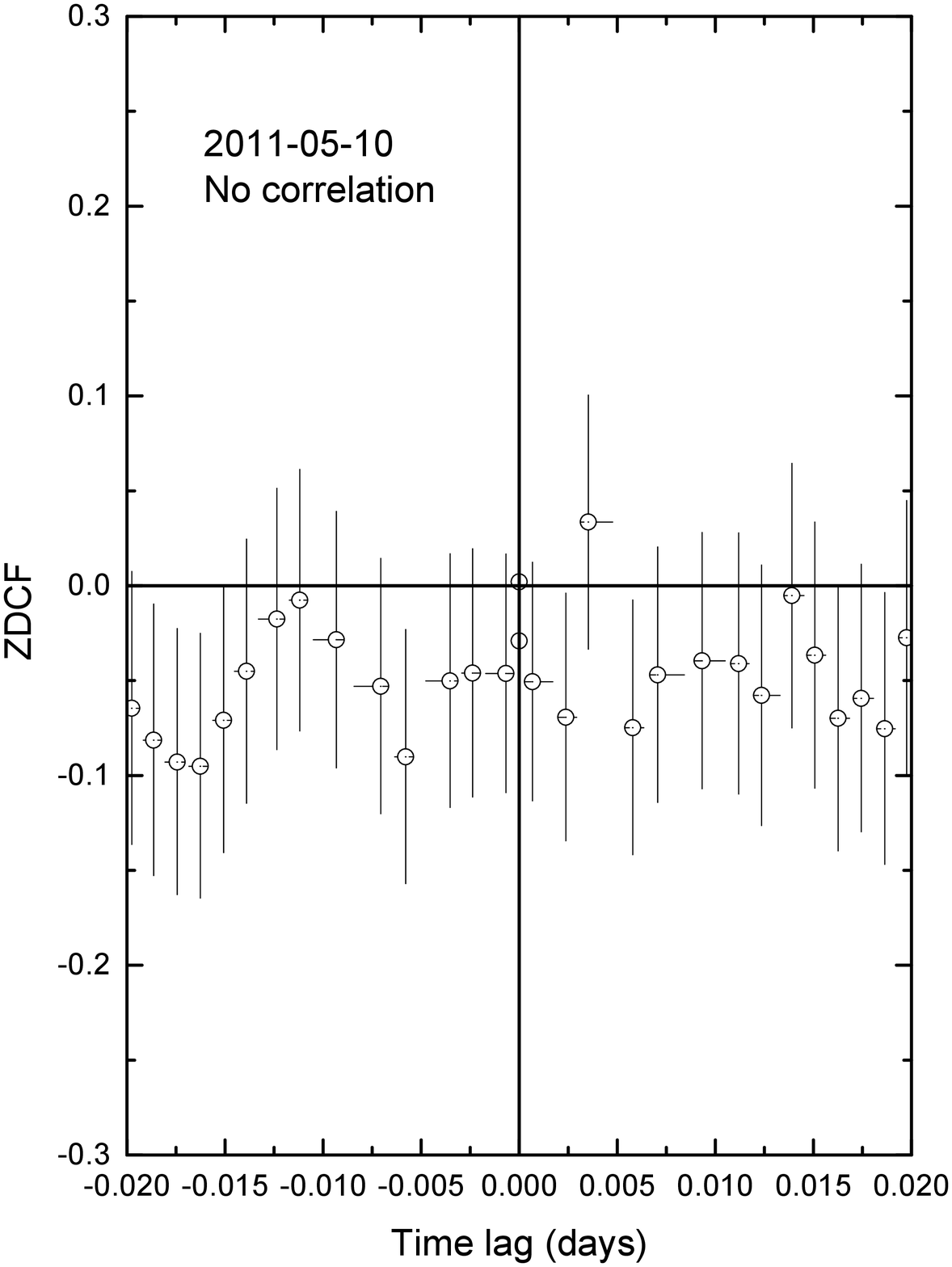}
  \includegraphics[angle=-90,scale=0.25]{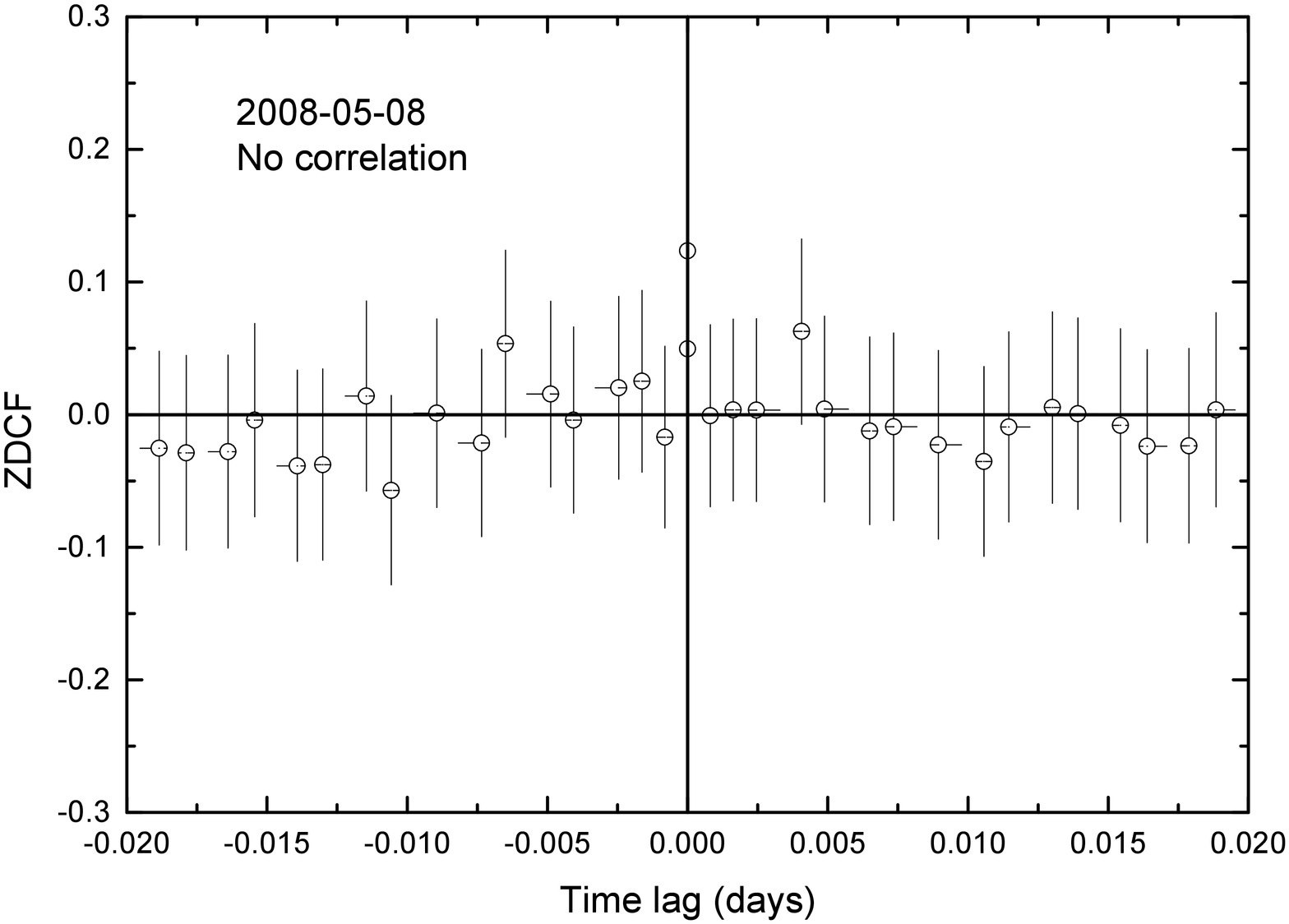}
  \includegraphics[angle=-90,scale=0.25]{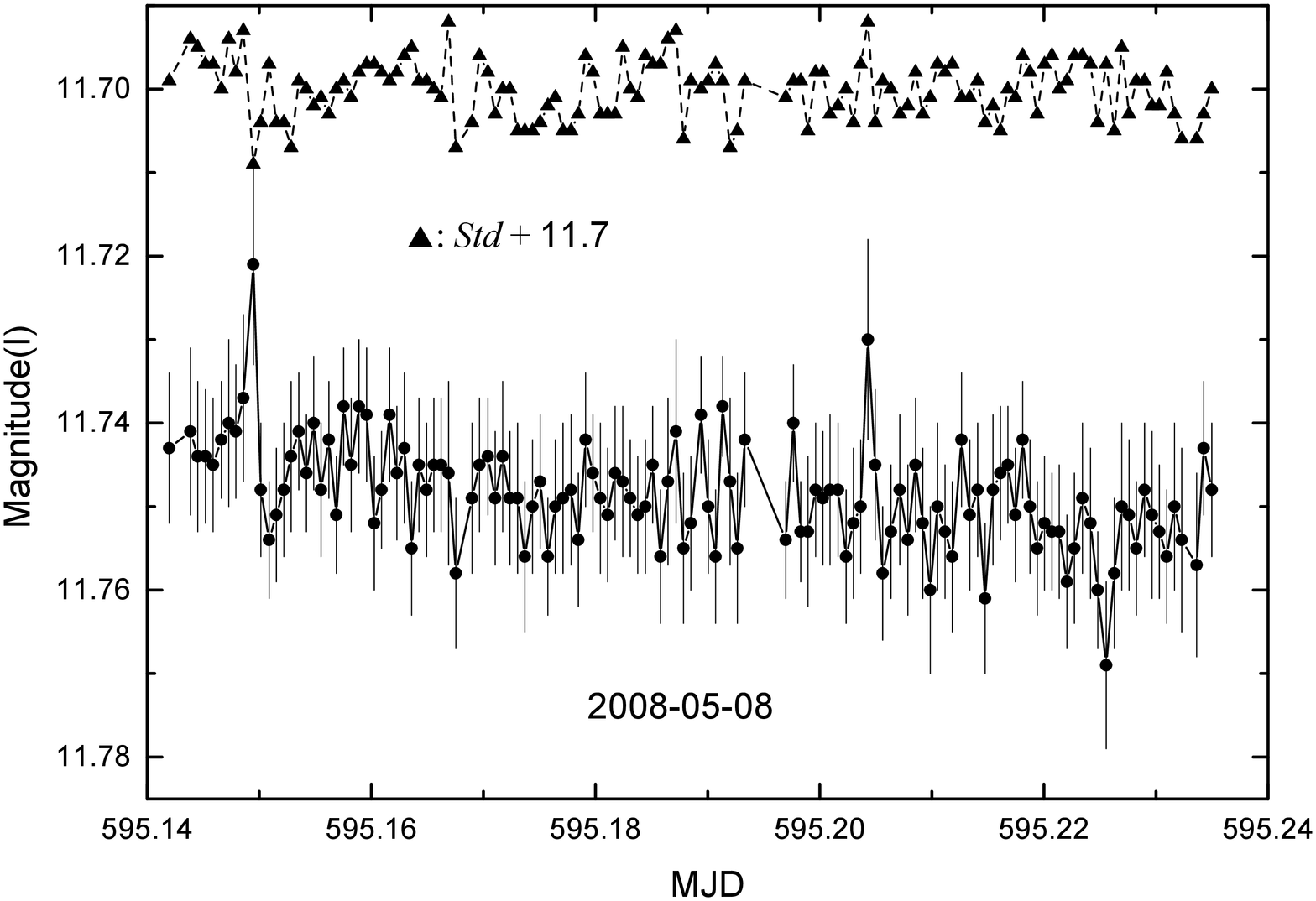}
 \end{center}
 \caption{ZDCFs calculated from the LCs in Figure 4 for 3C 273 and $Std^{\ast}$. The last two panels are the ZDCF
  and the LCs on 2008 May 08, and the LCs have the same symbols as Figure 2.}
  \label{fig6}
\end{figure*}


\begin{figure*}
 \begin{center}
  \includegraphics[angle=-90,scale=0.40]{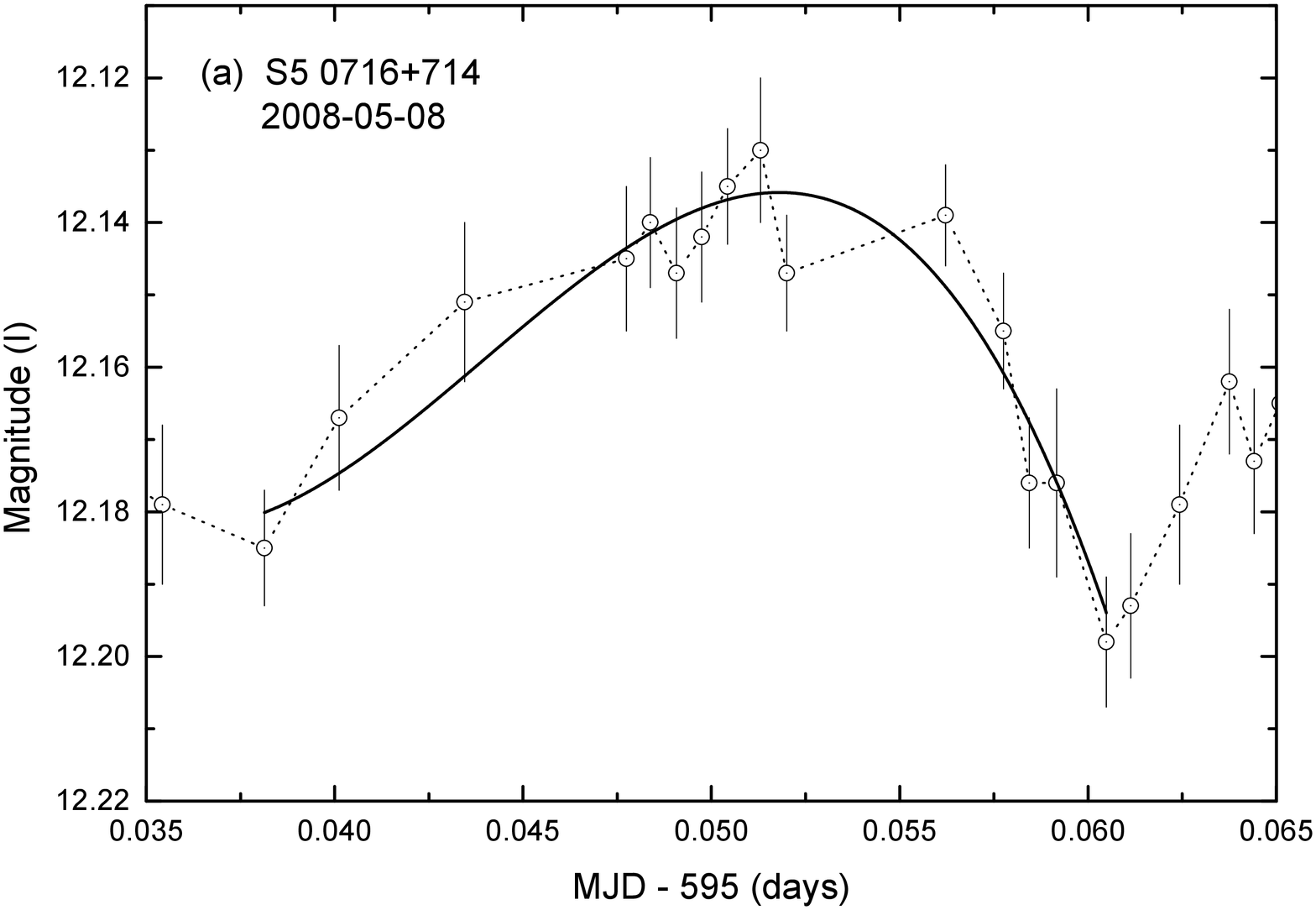}
  \includegraphics[angle=-90,scale=0.40]{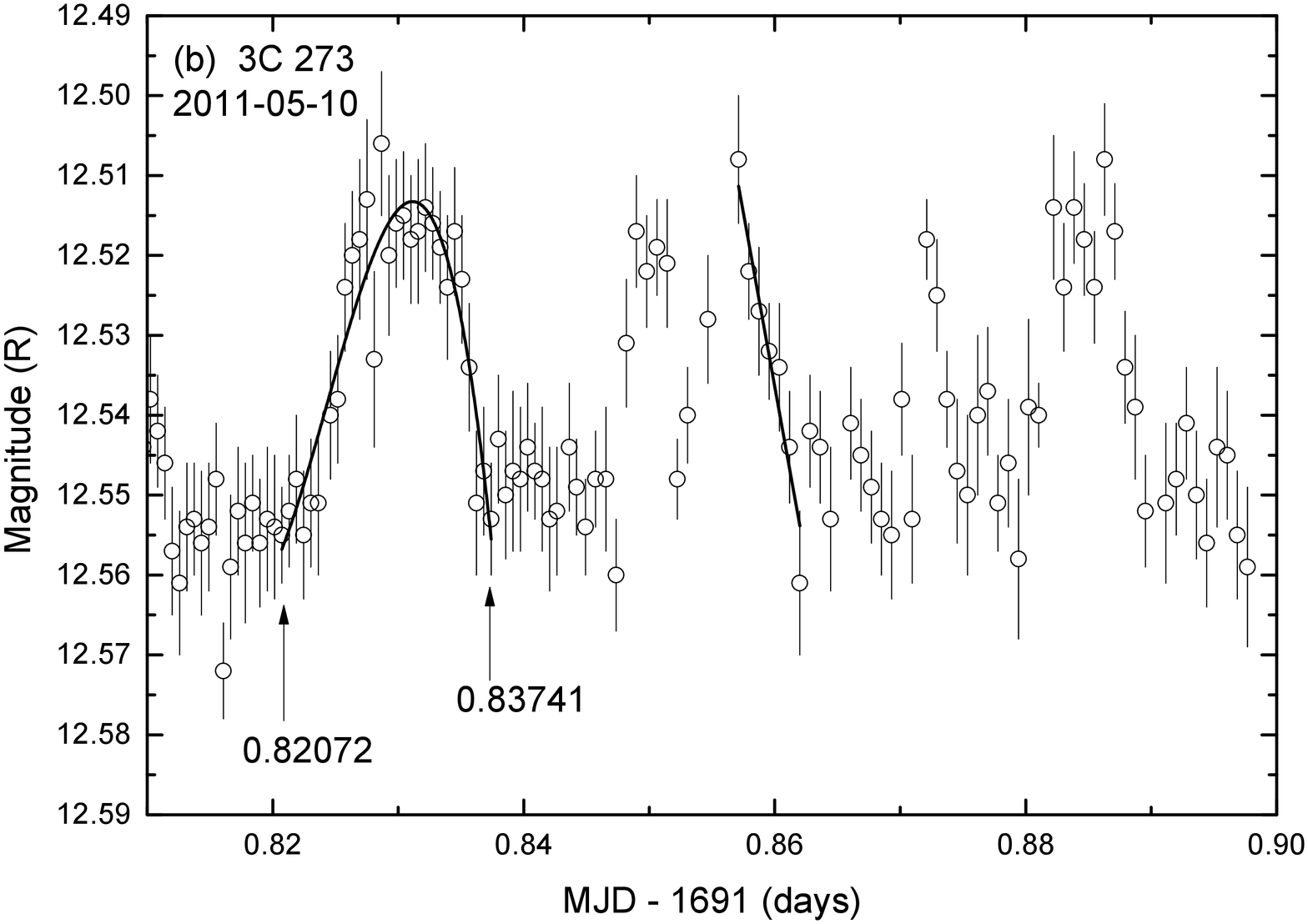}
 \end{center}
 \caption{Zoomed-in the IDV LCs of 3C 273 and S5 0716+714. Y-axes denote the apparent magnitudes. (a) The solid curve is the best third-order polynomial fitting to the 15 data points of the flare with the $y$ errors. (b) The straight line is the best linear fitting to the 7 data points with the $y$ errors. The numbers in panel are the corresponding times of data points denoted by the arrows, giving the duration of the flare. The solid curve is the best third-order polynomial fitting to the 29 data points of the flare with the $y$ errors.}
  \label{fig7}
\end{figure*}

\begin{figure*}
  \begin{center}
    \includegraphics[angle=-90,scale=0.40]{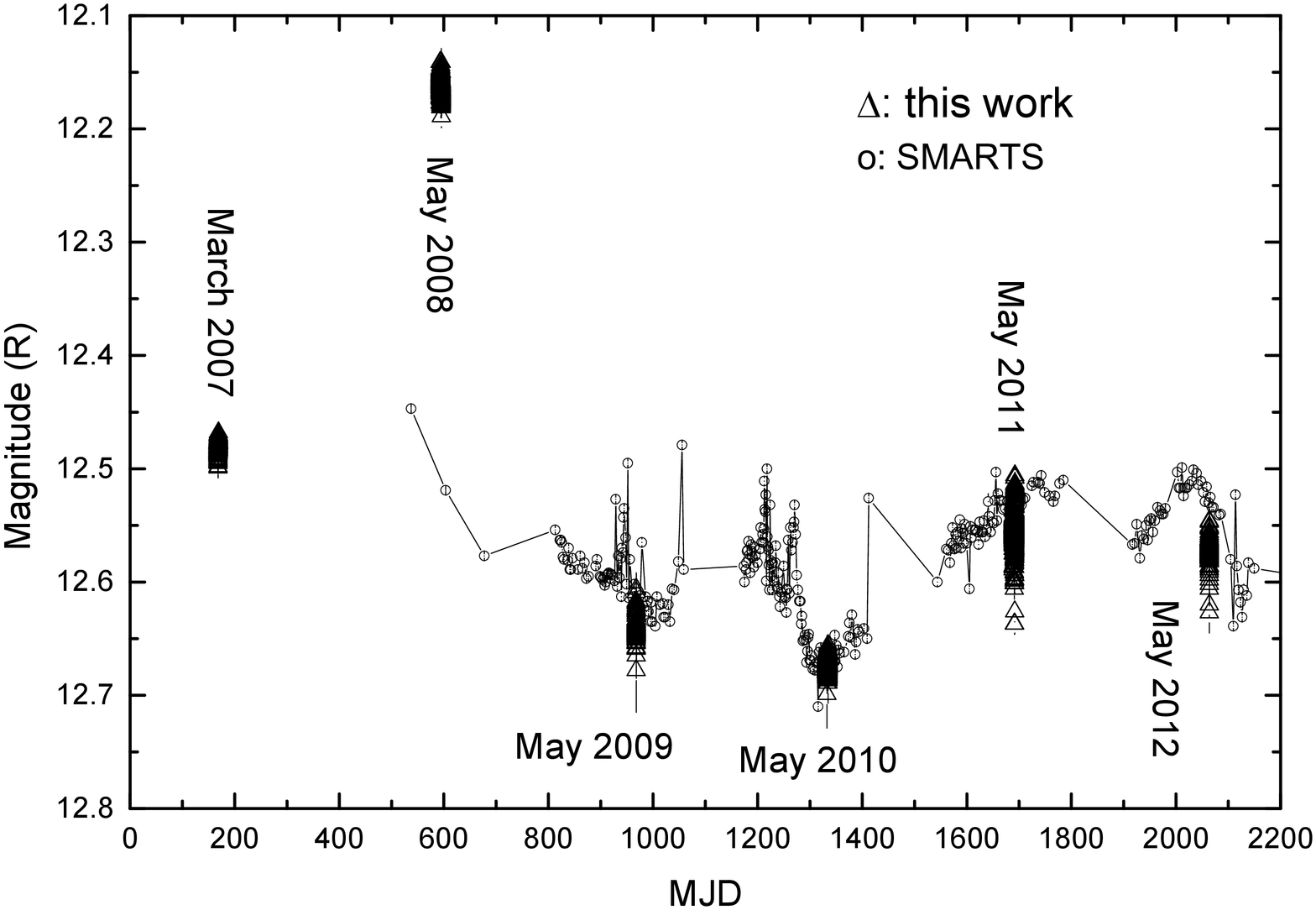}
  \end{center}
  \caption{Our long term LCs of 3C 273, and those of SMARTS \citep{Bo12}. Y-axis denotes the apparent magnitude. The $I$ magnitudes in 2007 March, 2008 May, and 2009 May are converted into the $R$ magnitudes by plus 0.42 mag, derived from the color index derived from \citet{Xi17}.}
  \label{fig8}
\end{figure*}


\end{document}